\documentclass[ aip, amsmath,amssymb, reprint,]{revtex4-1}

\usepackage{amsmath}
\usepackage{graphicx}
\usepackage{multirow}
\usepackage{array}
\usepackage{color} 
\usepackage{booktabs}
\usepackage{amsfonts}
\usepackage{hyperref}
\usepackage{textcomp}
\usepackage{ulem}
\usepackage[T1]{fontenc}
\usepackage{mathptmx}

\begin{document}

\title{A millikelvin scanning tunneling microscope in ultra-high vacuum with adiabatic demagnetization refrigeration}

\author{Taner Esat}
\affiliation{Peter Gr\"unberg Institute (PGI-3), Forschungszentrum J\"ulich, 52425 J\"ulich, Germany}
\affiliation{J\"ulich Aachen Research Alliance (JARA), Fundamentals of Future Information Technology, 52425 J\"ulich, Germany}
\author{Peter Borgens}
\altaffiliation{Current address: Peter Gr\"unberg Institute (Cryo-Lab), Forschungszentrum J\"ulich, 52425 J\"ulich, Germany}
\affiliation{Peter Gr\"unberg Institute (PGI-3), Forschungszentrum J\"ulich, 52425 J\"ulich, Germany}
\affiliation{Experimentalphysik IV A, RWTH Aachen University, 52074 Aachen, Germany}
\author{Xiaosheng Yang}
\affiliation{Peter Gr\"unberg Institute (PGI-3), Forschungszentrum J\"ulich, 52425 J\"ulich, Germany}
\affiliation{Experimentalphysik IV A, RWTH Aachen University, 52074 Aachen, Germany}
\author{Peter Coenen}
\affiliation{Peter Gr\"unberg Institute (PGI-3), Forschungszentrum J\"ulich, 52425 J\"ulich, Germany}
\affiliation{mProbes GmbH, 52428 J\"ulich, Germany}
\author{Vasily Cherepanov}
\affiliation{Peter Gr\"unberg Institute (PGI-3), Forschungszentrum J\"ulich, 52425 J\"ulich, Germany}
\affiliation{mProbes GmbH, 52428 J\"ulich, Germany}
\author{Andrea Raccanelli}
\altaffiliation{Current address: Peter Gr\"unberg Institute (Cryo-Lab), Forschungszentrum J\"ulich, 52425 J\"ulich, Germany}
\affiliation{Cryovac GmbH \& Co KG, 53842 Troisdorf, Germany}
\author{F. Stefan Tautz}
\affiliation{Peter Gr\"unberg Institute (PGI-3), Forschungszentrum J\"ulich, 52425 J\"ulich, Germany}
\affiliation{J\"ulich Aachen Research Alliance (JARA), Fundamentals of Future Information Technology, 52425 J\"ulich, Germany}
\affiliation{Experimentalphysik IV A, RWTH Aachen University, 52074 Aachen, Germany}
\author{Ruslan Temirov}
\email[Corresponding author: ]{r.temirov@fz-juelich.de}
\affiliation{Peter Gr\"unberg Institute (PGI-3), Forschungszentrum J\"ulich, 52425 J\"ulich, Germany}
\affiliation{II. Physikalisches Institut, Universit\"at zu K\"oln, 50937 K\"oln, Germany}

\date{\today}

\begin{abstract}
We present the design and performance of an ultra-high vacuum (UHV) scanning tunneling microscope (STM) that uses adiabatic demagnetization of electron magnetic moments for controlling its operating temperature in the range between 30 mK and 1 K with the accuracy of up to 7 $\mu$K. The time available for STM experiments at 50 mK is longer than 20 h, at 100 mK about 40 h. The single-shot adiabatic demagnetization refrigerator (ADR) can be regenerated automatically within 7 hours while keeping the STM temperature below 5 K. The whole setup is located in a vibrationally isolated, electromagnetically shielded laboratory with no mechanical pumping lines penetrating through its isolation walls. The 1K pot of the ADR cryostat can be operated silently for more than 20 days in a single-shot mode using a custom-built high-capacity cryopump. A high degree of vibrational decoupling together with the use of a specially-designed minimalistic STM head provides an outstanding mechanical stability, demonstrated by the tunneling current noise, STM imaging, and scanning tunneling spectroscopy measurements all performed on atomically clean Al(100) surface.
\end{abstract}

\maketitle

\section{Introduction}

Scanning tunneling microscopy (STM) performed in ultra-high vacuum (UHV) at millikelvin (mK) temperatures enables studies of delicate quantum phenomena that occur on surfaces \cite{Niimi2009,Song2010_1,Levy2013,Eltschka2014,Roychowdhury2015,Sun2015,Jack2015,Ast2016,Jack2016,Natterer2016,Hamidian2016,Feldman2017,Clark2018,Machida2019,Senkpiel2020,Weerdenburg2020,Nuckolls2020,Steinbrecher2021}. The growing interest in phenomena like quantum entanglement will most probably boost the demand for mK STM instrumentation. In contrast to STMs operating at higher temperatures, the mK STM technology is much less common. Virtually all of the existing mK STMs share one common feature: They use the technique of $^{3}$He-$^{4}$He dilution refrigeration (DR) to reach base temperatures below 100 mK \cite{Moussy2001,leSueur2006,Kambara2007,Song2010,Singh2013,Assig2013,Misra2013,Roychowdhury2014,vonAllworden2018,Machida2018,Balashov2018,Wong2020,Schwenk2020}. DR is a powerful technique which combines well with high magnetic fields and UHV conditions, but it also has its shortcomings. Most critically for the STM operation, DR continuously circulates cryoliquids in the microscope's vicinity, leading to a higher level of mechanical noise. In these circumstances, the search for alternative implementations of mK STM seems to be well justified. 

Here we describe the design and performance of the first-ever mK STM cooled by adiabatic demagnetization refrigeration (ADR) of electronic spins \cite{Pobell2007}. Although ADR was the first technique to reach temperatures below 1 K, it has never been applied to STM, presumably, due to the low thermal stability of inorganic salts typically used for ADR. Despite this, our results show that an ADR-based mK STM operating under UHV conditions is feasible, and, in some respects, perhaps an even more attractive alternative to the existing approaches. In particular, the use of ADR provides several significant advantages: First, it enables operation in a mechanically quiet environment. Second, the solid-state character of ADR-based devices makes the mK STM design very modular, which simplifies its everyday operation and makes the process of its further development much more efficient. Third, ADR enables accurate and fast control of the STM temperature in a wide range without using additional heaters.

\section{System layout}
\label{section: Layout}

\begin{figure*}
\centering
\includegraphics [angle=0, width=16.8 cm]{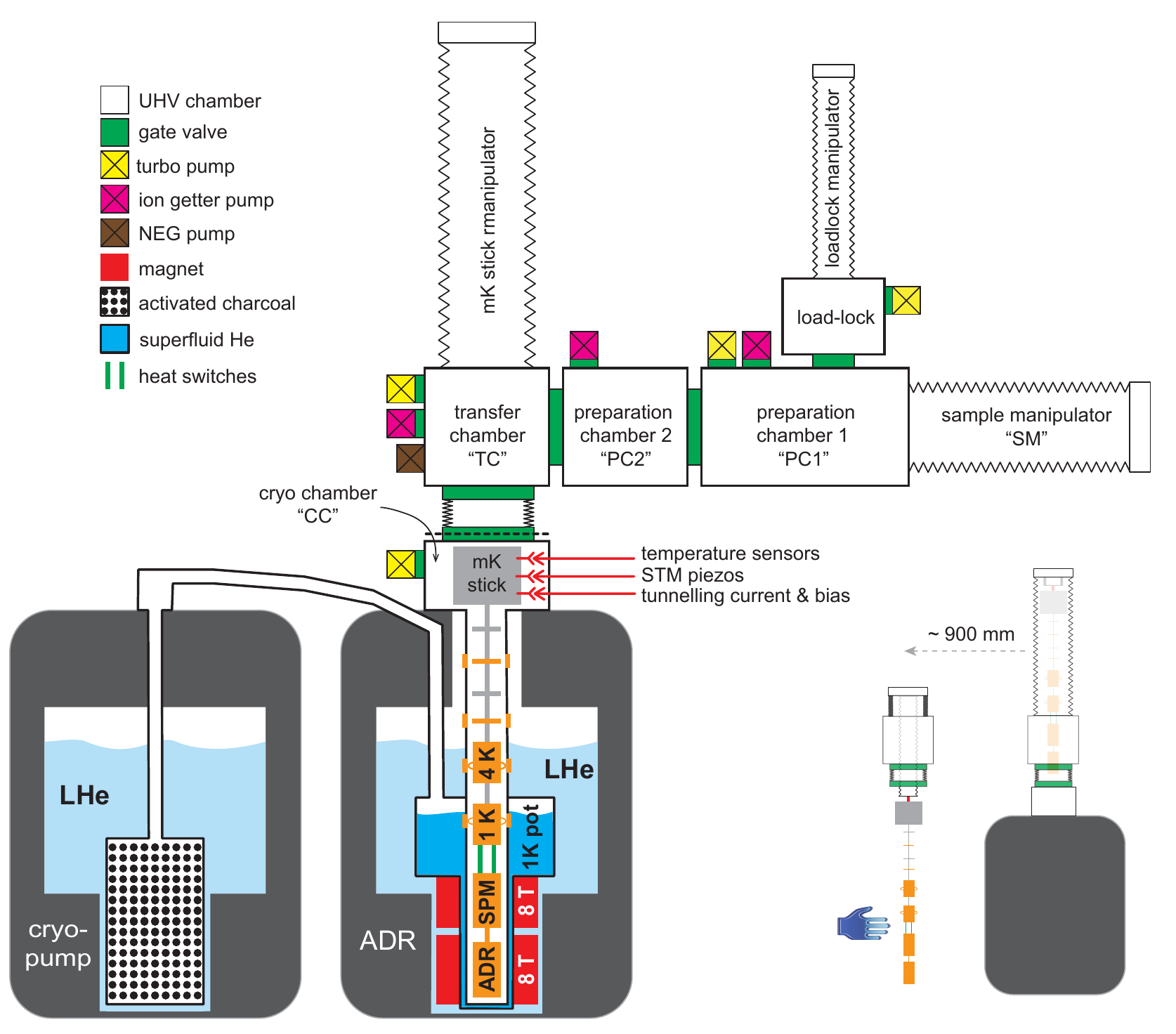}
\caption []{Schematic layout of the mK STM setup comprising the UHV chambers, the ADR cryostat hosting the mK stick, and the high-capacity cryopump. The main UHV system, which includes the load-lock, preparation chambers 1 \& 2, and transfer chamber connects to the cryostat via a flexible bellow. To extract the mK stick out of the vacuum, the cryostat and the UHV system must be separated at the plane marked by the dashed line. \textit{Bottom right:} The inset shows the extraction of the mK stick out of UHV. The frame supporting the UHV system is translated sideways in the direction perpendicular to the plane of the main figure for the extraction.} 
\label{fig: Layout}
\end{figure*}

We start presenting our UHV ADR mK STM by describing the system's layout. The aim is to provide a general understanding of the system functionalities rather than to present full details of its technical design. The setup comprises a UHV system, two superinsulated 300 liter liquid helium (LHe) dewars, and a removable millikelvin insert (mK stick) on which the STM is mounted (see Fig.~\ref{fig: Layout}). One of the LHe dewars hosts the ADR cryostat and the mK stick, while the other one houses a high-capacity cryopump that operates the 1K pot of the ADR cryostat during the silent regime of the mK STM operation. The whole setup fits inside a vibrationally isolated and electromagnetically shielded laboratory \cite{Voigtlander2017} with an area of 5 $\times$ 4 m$^2$ and a height of 4 m.

\subsection{UHV chamber}
\label{section: UHV-chamber}

The UHV chamber consists of five sections separated by gate valves, as Fig.~\ref{fig: Layout} shows. Below we give a brief description of each section. Preparation chamber 1 (PC1) serves for the UHV sample preparation. It is equipped with a UHV sample manipulator (SM) manufactured by VAb \cite{VAB}. The SM is used to prepare samples and to transfer them to the STM. The equipment of the PC1 also includes a focused ion gun from Focus GmbH \cite{Focus} and AES-LEED optics from SPECS \cite{Specs}. The preparation chamber 2 (PC2) is meant to host more preparation techniques in the future, but at the moment, it is not in use. The base pressure in the PC1 and the PC2 is typically around $2\times 10^{-10}$ mbar.

The cryochamber (CC) that hosts the mK stick during STM experiments as shown in Fig.~\ref{fig: Layout} comprises the ADR cryostat bore and a small chamber inside which the mK stick head is locked and which is positioned directly on top of the cryostat. The CC has three linear motion feedthroughs equipped with pogo-pin contact plates, which establish electrical connections to the mK stick by pressing against the corresponding contact plate of the mK stick head (cf. Fig.~\ref{fig: mK-stick}). The low-voltage wiring necessary for the temperature sensors' readout and control of the heat-switches passes through a 51-pin micro SUB-D UHV feedthrough from VACOM \cite{VACOM}. The high-voltage cabling of the STM piezo passes through a separate 15-pin SUB-D UHV feedthrough. Finally, the tunneling current and the bias wires pass through a 4$\times$SMA coaxial feedthrough purchased from Allectra \cite{Allectra}. The chamber has two additional linear motion stages for operating the mechanisms, which lock (unlock) the mK stick to (from) the manipulator or the cryostat.

The transfer chamber (TC) hosts a custom-designed UHV manipulator for lifting and loading the mK stick (section \ref{section: mK-stick manipulation}). In addition to the ion and turbopump, the TC is equipped with a non-evaporable getter (NEG) pump from SAES \cite{SAES}. The TC connects to the CC through a 50 mm long flexible DN100 UHV bellow with a gate valve attached on each side. The mechanical connection between the TC and the CC can be interrupted without disrupting the STM experiment. Having such an option could in future be useful for minimizing electrical or vibrational noise further. More importantly, disconnecting the TC from the CC makes it possible to remove the mK stick out of the UHV without warming-up the cryostat. Because the current version of the mK stick is not bakeable (section \ref{section: mK-stick}), the pressure in the TC is at the level of $3\times 10^{-9}$ mbar.

\subsection{ADR cryostat} 
\label{section: ADR-cryostat}

The superinsulated LHe bath cryostat that cools the mK stick to 1 K has been designed and manufactured by Cryovac \cite{Cryovac}. The boil-off rate of LHe from the dewar of the cryostat is about 0.85 l of LHe per hour. This relatively high boil-off rate is mainly a consequence of the wide neck of the cryostat, necessary for the magnet setup, compared to its relatively short height which was intentionally minimized to fit the whole setup in the 4 m high laboratory space. The four retractable evaporation-cooled current leads of the magnets do not contribute to the LHe losses noticeably. The maximum holding time of the LHe cryostat reaches 14 days. Typically, it is refilled every 7--10 days. 

The system of superconducting magnets installed in the ADR cryostat was custom-designed by Cryomagnetics \cite{Cryomagnetics}. It consists of a pair of superconducting coils stacked axially on top of each other, as shown schematically in Fig.~\ref{fig: Layout}. The lower coil with a maximum field of 8 T performs the ADR, while the upper 8 T coil generates the $B$ field in the sample region. The ADR magnet provides the necessary field homogeneity in the ADR pill region. The sample magnet has a set of compensating coils that reduce its stray field in the ADR magnet region below 5 mT at the maximum field of 8 T.

The inner bore of the cryostat, comprises a 50 mm wide stainless steel tube, which at the same time is the inner wall of the CC, interrupted at specific heights by four gold plated copper rings (see Fig.~\ref{fig: Layout}). These copper rings serve as thermal coupling of the mK stick to the cryostat. The upper two rings contact the two mK stick copper baffles (see Fig.~\ref{fig: mK-stick}). The lower two rings thermalize the 4K and 1K stages, respectively, of the mK stick.
 
The lowest copper ring thermalizing the 1K stage is a part of the outer wall of the 1K pot. The 1K pot with a volume of 1.3 liter receives LHe from the dewar of the bath cryostat via a capillary equipped with a needle valve for the regulation of the LHe flow. The 1K pot is thermally isolated by the surrounding inner vacuum chamber (IVC) (not shown in Fig.~\ref{fig: Layout} to avoid clutter). When the supply capillary is closed, the 1K pot enters the single-shot mode, reaching a minimal temperature of 0.975 K. The 1K pot was designed to operate at least two weeks under a load of up to 3 mW. In practice, the operation time of the 1K pot is about 25 days in the single-shot mode without interruption, indicating a much smaller thermal load in our setup.
  
\subsection{High-capacity sorption pump} 

The high-capacity cryopump was also designed and manufactured by Cryovac \cite{Cryovac} and uses activated charcoal as the sorption material \cite{Haefer1989}. Sorption pumps have been extensively used for $^{3}$He and $^{4}$He cryogenics for more than half a century \cite{Shvets1966}, allowing miniaturization of sub-Kelvin coolers \cite{Torre1985}. The capacity of our sorption pump as well as its external mode of operation makes it suitable for any pumped Helium stage that is usually operated with a mechanical pump, without any modification of the latter. The cryopump capacity was designed to provide more than two weeks of uninterrupted operation of the 1K pot of the ADR cryostat under a heat load of 3 mW. It operates steadily for about a month in real conditions, after which it can be fully regenerated within three hours. The sorption pump is surrounded by an evacuated volume (not shown in Fig.~\ref{fig: Layout} to avoid clutter) that can be filled with LHe from the main LHe bath through a needle valve. When evacuated, this volume provides thermal insulation from the LHe bath and allows the regeneration of the pump by heating it to 40 K and pumping the desorbing helium gas with a scroll pump. The boil-off rate of the superinsulated LHe dewar of the cryopump is about 0.5 liter LHe per hour, which results in a maximum holding time of three weeks. We typically refill the cryopump dewar once in 14 -- 20 days.

\subsection{mK stick} 
\label{section: mK-stick}

\begin{figure}
\includegraphics [angle=0, width=7.9 cm]{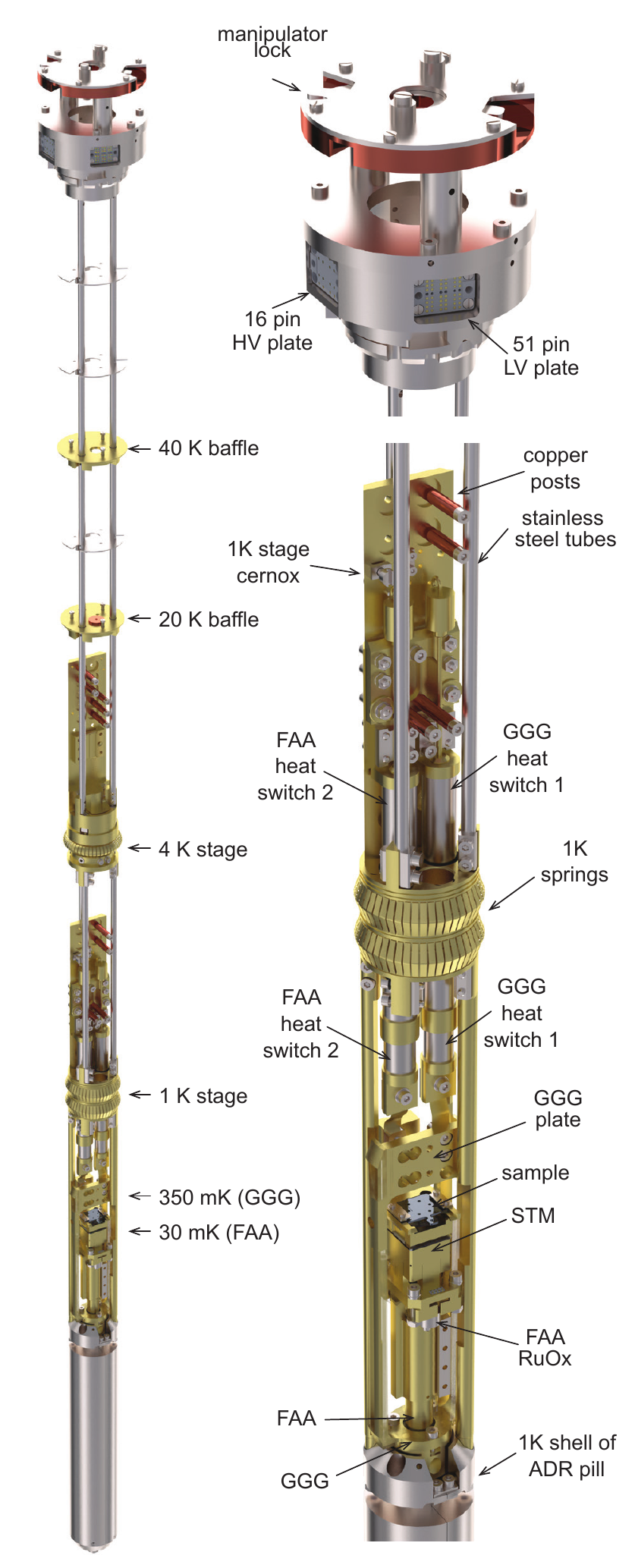}
\caption []{Rendered CAD model of the mK stick. \textit{Left:} mK stick in its full length of $156.5$ cm. The arrows indicate locations of different temperature stages. \textit{Top right:} Head of the mK stick with the mechanism locking it to the vertical manipulator that loads it into the cryostat. Two of the contact plates used for establishing the electrical contacts to the temperature sensors and the STM piezo are also visible. The third contact plate establishing the coaxial bias and tunneling current contacts is located on the back side. \textit{Bottom right:} Image detail of the mK stick below the 4K stage without wiring. For further information, see text.} 
\label{fig: mK-stick}
\end{figure}

The mK stick is a crucial part of the mK STM, as it hosts both the STM and the paramagnetic pill (ADR pill) necessary for reaching millikelvin temperatures. As noted above, the mK stick has no permanent electrical or mechanical connections to the rest of the setup. Therefore, it can be quickly extracted out of UHV and even exchanged without warming up the ADR cryostat. This feature allows for a unique degree of modularity of our setup.

The mK stick comprises the following essential parts (cf. Fig.~\ref{fig: mK-stick}): the head featuring electrical contact plates and the locking mechanism for attaching the mK stick either to the manipulator or to the top of the cryostat; three thin-walled stainless steel tubes forming the structural backbone of the mK stick down to its 1K stage, below which the structure is supported by three gold plated copper rods; three stainless-steel baffles for radiation protection; two copper baffles thermalized to the 40 K and 20 K copper rings of the CC (section \ref{section: ADR-cryostat}); the 4K stage with the gold plated CuBe spring establishing the thermal contact to the 4K copper ring of the CC; the 1K stage with the two gold plated CuBe springs coupling it thermally to the 1K pot; two $^{3}$He gas-gap heat switches, designed and manufactured by Chase Research Cryogenics \cite{ChaseCryogenics}; a home-built STM; and finally a two-stage ADR pill, designed and manufactured by Entropy \cite{Entropy} and attached to the very bottom of the mK stick. 

For monitoring the mK stick temperature, we use four temperature sensors: Two calibrated Cernox sensors (Lakeshore) \cite{Lakeshore} are mounted at the 1K and 4K stages, respectively. Two calibrated RuOx sensors from Entropy \cite{Entropy} monitor the temperature of the two stages of the ADR pill (see below). The sensor wiring running between 300 K and 1 K is made of a twelve-twisted-pair constantan ribbon from CMR \cite{CMR}. Below the 1K stage, the wiring continues with three individual four-twisted-pair NbTi ribbons acquired from CMR. 

For the wiring of the STM, we use a shielded twisted-pair cable obtained from GVL Cryogenics \cite{GVL}. This cable has a braided constantan shield with a resistance of 6 $\Omega$/m, while the 0.1 mm diameter inner brass conductors have a resistance 8 $\Omega$/m. The inner conductors are varnished and covered with Teflon for additional electric isolation. We have tested the cables in UHV and detected no appreciable outgassing at the highest tested temperature of 150 \textdegree{}C.

Due to its construction, the piezo system of our home-built STM (section \ref{section: STM}) needs only five electrical connections for its operation. These are implemented with three shielded twisted-pair lines. The wiring of the STM bias and the tunneling current consists of two twisted-pair lines in which we turned the two inner conductors of the pair into one by soldering them together at both ends. All twisted-pair cables run from 300 K down to 30 mK without physical interruption. Each cable is wrapped around two copper posts for thermalization: one located at the 4K and the other at the 1K stage (Fig.~\ref{fig: mK-stick}). Furthermore, we thermalize these cables by pushing them into 5 to 15 cm long, 3 mm deep, and 0.8 mm wide groves in the copper bodies of both the 300 mK and the 30 mK plates of the mK stick. The last few centimeters of the bias and the tunneling current wires consist of unshielded NbTi superconducting wire to improve thermal decoupling and additionally block high-frequency noise \cite{vonAllworden2018}.

\begin{figure}
\includegraphics [angle=0, width=7.9 cm]{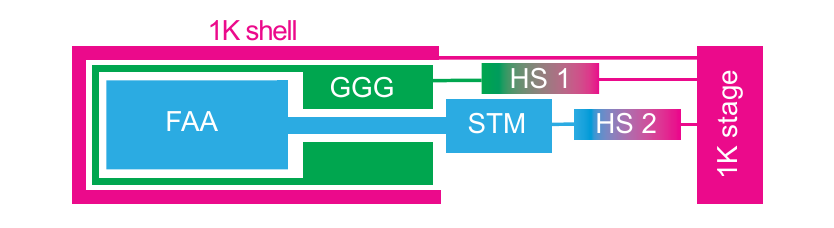}
\caption []{Principal scheme of thermal flow in the ADR pill. Two gas-gap heat switches connect the FAA and GGG stages to the 1K stage. In the ``OFF'' state of the heat switches, the FAA and GGG stages decouple thermally from 1K stage. The FAA and GGG stages are also thermally isolated from each other. The STM is firmly screwed to the FAA stage. The 1K shell of the ADR pill is permanently well-connected to the 1K stage of the mK stick.} 
\label{fig: ADR-pill}
\end{figure}

The bottom of the mK stick features a commercially acquired ADR pill that contains two paramagnetic materials: Ferric ammonium alum (FAA) with the chemical formula NH$_{4}$Fe(SO$_{4}$)$_2 \cdot 12 $H$_{2}$O, and gadolinium gallium garnet (GGG)---Gd$_{3}$Ga$_{5}$O$_{12}$. Fig.~\ref{fig: ADR-pill} reveals the principal scheme of the ADR pill, which consists of three thermally decoupled stages. The outer aluminum shell is in good thermal contact with the 1K stage of the mK stick. It encloses the second shell made of gold-plated copper with a 201 g single crystal of GGG firmly attached to it. This assembly that we refer to as the GGG stage reaches 350 mK during the ADR cooling cycle (section \ref{section: ADR cycle}). As mentioned above, the GGG stage provides precooling for the wiring that passes to the STM and simultaneously reduces the heat leak from the 1K stage towards the ADR pill lowest temperature, FAA stage. The GGG stage is also attached to one of the two gas-gap heat switches (cf. Fig.~\ref{fig: mK-stick} and section \ref{section: ADR cycle}) that connects it to the 1K stage of the mK stick. A calibrated RuOx sensor fixed to the GGG stage provides the reading referred to as ``GGG temperature'' or $T_\mathrm{GGG}$.

The core of the ADR pill -- the FAA stage -- that reaches the lowest temperature of 26~mK comprises a single crystal of FAA with a mass of 210 g enclosed in a UHV-tight stainless steel container. The FAA crystal is in good thermal contact with a thick copper rod that sticks out of the container and provides thermalization for the STM and the second RuOx sensor, the reading of which is then referred to as FAA temperature or $T_\mathrm{FAA}$. The FAA stage of the ADR pill is also connected to the 1K stage of the mK stick through its own gas-gap heat switch (cf. Fig.~\ref{fig: mK-stick} and section \ref{section: ADR cycle}).

One final note on the temperature stability of the ADR pill is due. Unlike the GGG crystal that is thermally stable, the FAA crystal contains water and its melting temperature is about 40 \textdegree{}C \cite{Wilson1999}. This makes the ADR pill and the whole mK stick not-bakeable. As will be demonstrated below, this technical complication does not prevent us from preparing and working with atomically clean sample surfaces. At the same time, we are also developing a new type of bakeable ADR pill using novel, thermally-stable magnetocaloric materials \cite{Jang2015}.

\subsection{STM} 
\label{section: STM}

\begin{figure}
\centering
\includegraphics [angle=0, width=7.9 cm]{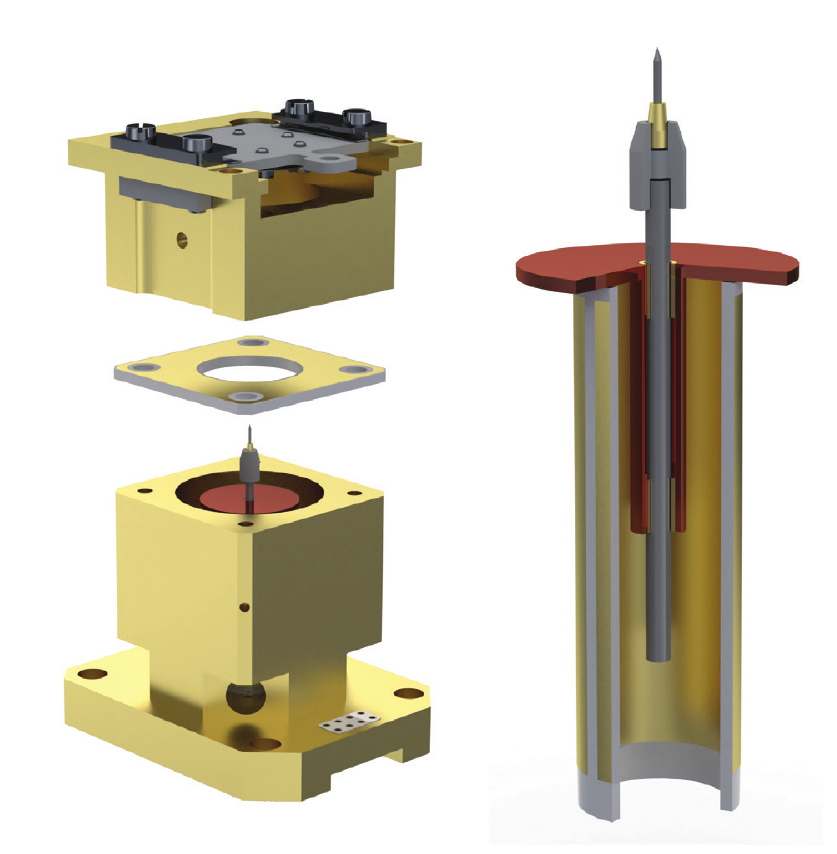}
\caption []{\textit{Left:} Exploded view of the home-built STM. The top part of the STM is electrically isolated from the STM body by a sapphire plate. The STM body hosts a single piezo tube used for both the coarse and fine motion of the STM tip. \textit{Right:} Cut view of the piezo tube, revealing the stick slip coarse motor (see text for details).}
\label{fig: STM}
\end{figure}

\begin{figure}
\centering
\includegraphics [angle=0, width=7.9 cm]{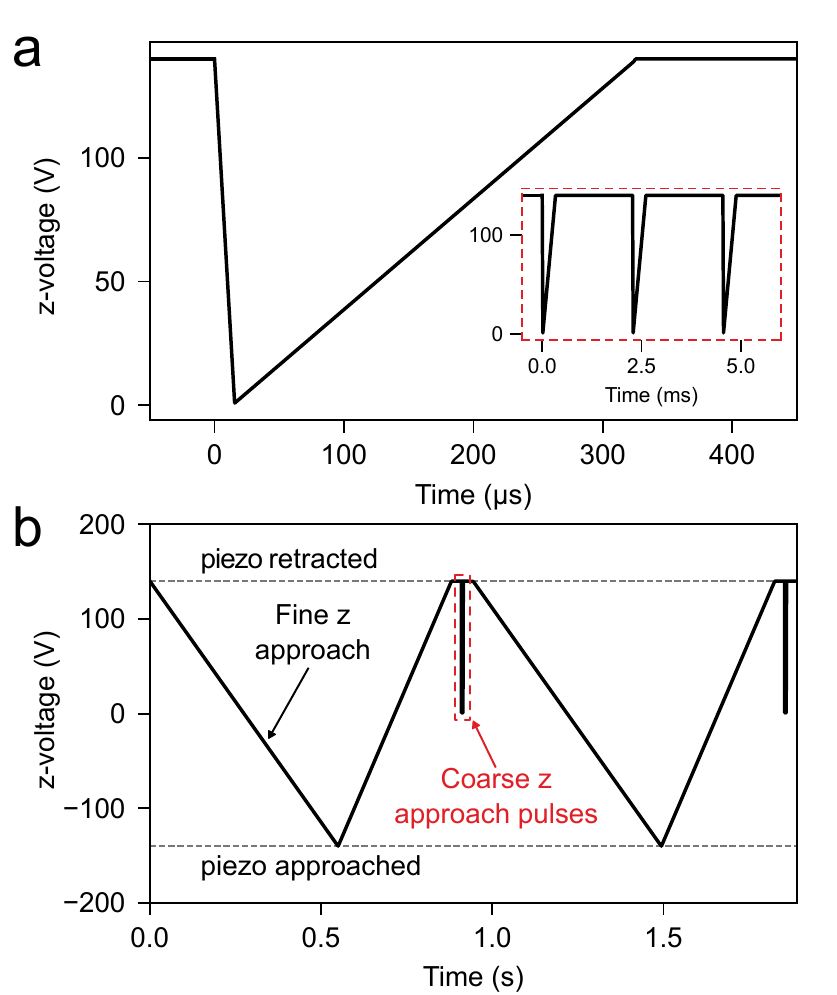}
\caption []{a) A voltage pulse producing a single coarse $z$ step towards the surface. When inverted in time, the pulse produces a single step away from the surface. \textit{Inset}: A sequence of three coarse $z$ pulses typically employed during auto-approach. b) A voltage pulse sequence applied during the auto-approach.}
\label{fig: STM-coarse}
\end{figure}

A unique feature of our home-built STM (cf. Fig.~\ref{fig: STM}) is that both $z$-coarse and $xyz$-scanning functions are implemented in a single piezoceramic tube. Such a design makes the STM very compact and thus less sensitive to mechanical noise. It also needs fewer high voltage lines for its operation, thus also reducing the thermal load on the mK stick. As shown in the right panel of Fig.~\ref{fig: STM}, the $z$-coarse approach motor of our STM exploits the slip-stick mechanism: A tungsten rod, held by two CuBe springs inside a CuBe tube, which is firmly attached to the upper end of the piezo tube, can be accelerated slowly in the direction of the long axis of the tube by extending (contracting) the piezo. The tungsten rod makes a coarse $z$ step when the extension (contraction) of the piezo tube reverses rapidly, causing a high acceleration of the rod that eventually overcomes its friction with the springs. The voltage pulse used to produce a single coarse step towards the surface is shown in Fig.~\ref{fig: STM-coarse}a. The pulse is applied to the inner contact of the piezo tube against all its four outer contacts. To move in the opposite direction, the time profile of the pulse is inverted. The coarse steps are made with the piezo in its fully contracted state (i.e., under high positive voltage), which is necessary for the subsequent test of the tunneling contact after the coarse step during auto-approach. The pulse sequence used for auto-approach is shown in Fig.~\ref{fig: STM-coarse}b. 

The STM fine scanner is operated conventionally by applying voltages of opposite polarity to the corresponding pairs of $x$ and $y$ electrodes. For fine scanning of $z$, the inner contact of the piezo tube is biased against all four outer contacts. Operating with voltages of up to 150~V, we can scan an area of 2.5 $\times$ 2.5 $\mu$m$^2$ at the lowest millikelvin temperatures. It should be possible to reach a 6.5 $\times$ 6.5 $\mu$m$^2$ scan range by applying voltages of up to 400 V. 

Our STM accepts standard flag-type sample plates. The sample is inserted into the top part of the STM, electrically isolated from the rest of the STM body by a 1 mm thick sapphire plate covered from both sides with a thin film of gold to improve thermal contact.

\section{System operation and characteristics}
\label{section: System operation and characteristics}

\subsection{Initial cooldown}
\label{section: cooldown}

\begin{figure}
\includegraphics [angle=0, width=7.9 cm]{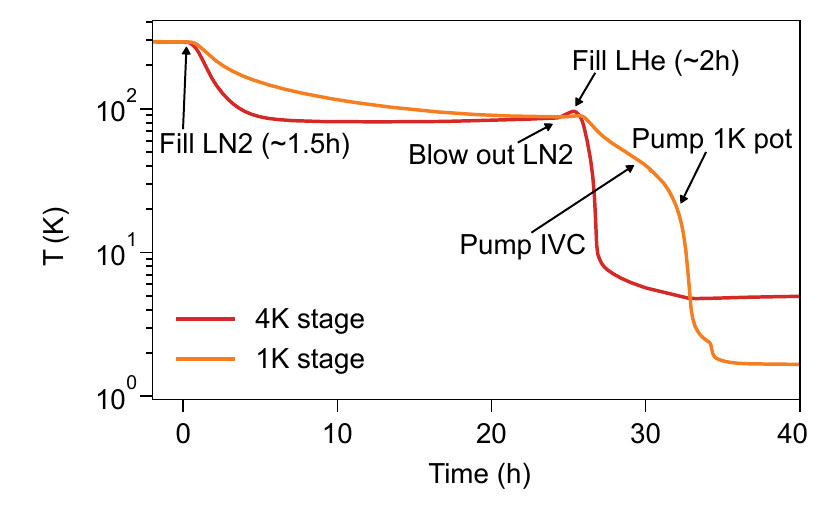}
\caption []{Evolution of the 4K stage and 1K stage temperatures of the mK stick during the initial cooldown from 300 K. For further information, see text.} 
\label{fig: Initial-cool}
\end{figure}

The initial cooldown of the ADR cryostat starts with precooling it to liquid nitrogen (LN2) temperature. Fig.~\ref{fig: Initial-cool} shows that the cooling of the mK stick from room temperature to 80 K takes about 24 hours. During this time, the IVC is kept at a pressure of few millibars of helium gas. After reaching the desired temperature and blowing LN2 out of the superinsulated dewar, the filling with LHe starts. When the temperature of the 1K stage reaches 40 K, the IVC has to be evacuated, and briefly afterwards the pumping of 1K pot with a mechanical pump starts. As Fig.~\ref{fig: Initial-cool} shows, after about 35 hours, the temperature of the 1K stage drops below 2 K. The cooldown of the cryopump proceeds analogously and is not described here in detail. The final temperatures of the 4K and 1K stages of the mK stick reached in different regimes of the ADR cryostat's operation are listed in Table \ref{tab: T}.
\\
\\
\begin{table}
\begin{center}
\begin{tabular}{ >{\centering\arraybackslash}p{2cm}|| >{\centering\arraybackslash}p{3cm}| >{\centering\arraybackslash}p{3cm}}
   & heat switches on & heat switches off\\
  \hline\hline
  4K Stage & 4.55 K to 5.10 K & 4.55 K to 5.10 K \\
  1K Stage & 1.70 K & 1.03 K \\
	GGG stage & 1.66 K & 1.02 K \\
	FAA stage& 1.68 K & 1.03 K \\ 
	1K pot & 1.04 K & 0.98 K
\end{tabular}
\caption{Temperatures of the different stages of the mK stick depending on the state of the heat switches. The 1K pot is operated in single-shot mode. Note that the temperature of the 4K stage varies with the LHe level of the ADR cryostat. Also note that turning the heat switches on raises the temperature of the FAA, GGG, and 1K stages, because the two 10 k$\Omega$ (room temperature value) resistors that are used to warm up the charcoal inside the heat switches (see section \ref{section: ADR cycle}) dissipate about 0.36 mW of power each when biased by 2 V. 
\label{tab: T}}
\end{center}
\end{table}

\subsection{mK stick manipulation}
\label{section: mK-stick manipulation}

\begin{figure}
\includegraphics [angle=0, width=7.9 cm]{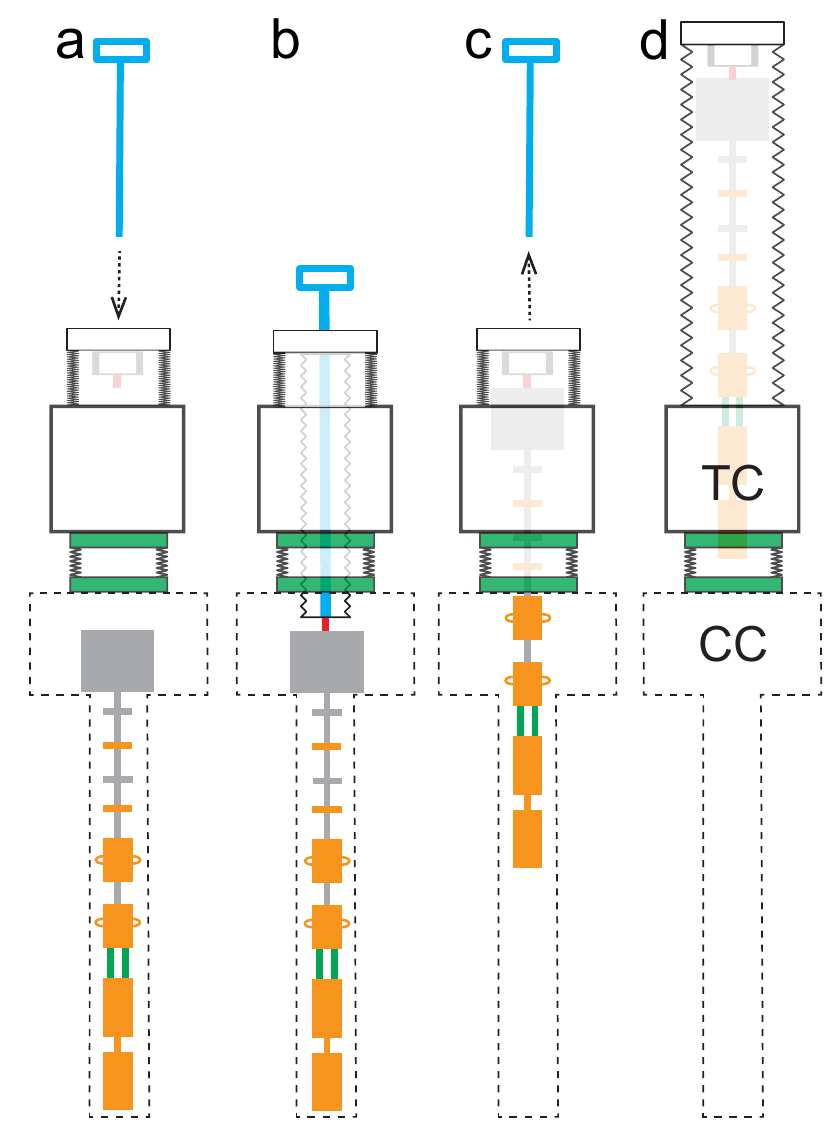}
\caption []{(a)-(d) The sequence of manipulation steps used to remove the mK stick from the ADR cryostat. To remove the mK stick from the vacuum, the manipulation continues with the sequence (d)-(a), executed after the vacuum connection between TC and CC is broken, TC is vented, and the UHV chamber frame shifted to the side by ca. 900 mm, which clears the space below the manipulator. The dashed line shows the contour of CC. For further information, see text and the inset of Fig.~\ref{fig: Layout}.} 
\label{fig: mK-stick-manipulation}
\end{figure}

Our mK STM is designed for full operation in laboratory rooms with a minimal height of only 4 m. This required the design and manufacture of a special manipulator for in-situ sample exchange and the extraction of the mK stick from the bath cryostat. The manipulator is motorized and operated with the help of a custom-made software. A typical mK stick manipulation cycle that is performed to load the sample into the STM (or remove it from the STM) is exhibited in Fig.~\ref{fig: mK-stick-manipulation}. 

In the initial position (Fig.~\ref{fig: mK-stick-manipulation}a), both the outer and inner bellows of the manipulator are in the contracted state and the stick resides in the STM measurement position (inside the CC) while all the electrical contacts to the mK stick are disengaged, i.e. all corresponding linear stages in the CC (Fig.~\ref{fig: Layout}) are in the retracted position. Next, a push-pull rod is attached to the bottom of the inner bellow (Fig.~\ref{fig: mK-stick-manipulation}a) and the bellow is expanded downwards until it reaches the head of the mK stick (Fig.~\ref{fig: mK-stick-manipulation}b). Then, the head of the mK stick is fixed to the manipulator by actuating the lock situated in the upper part of the mK stick (Fig.~\ref{fig: mK-stick}), and the manipulator lifts the mK stick (Fig.~\ref{fig: mK-stick-manipulation}c) by contracting the inner bellow to its original state. Here, the push-pull rod is removed, and the outer bellow of the manipulator is expanded, lifting the mK stick (Fig.~\ref{fig: mK-stick-manipulation}d) to the position at which the sample can be loaded into (or extracted from) the STM using the sample manipulator SM (see Fig.~\ref{fig: Layout} and section \ref{section: UHV-chamber}). The combined action of the outer and inner bellows allows for a vertical travel distance of the mK stick that exceeds the elongation of the manipulator itself. 

\begin{figure}
\centering
\includegraphics [angle=0, width=7.9 cm]{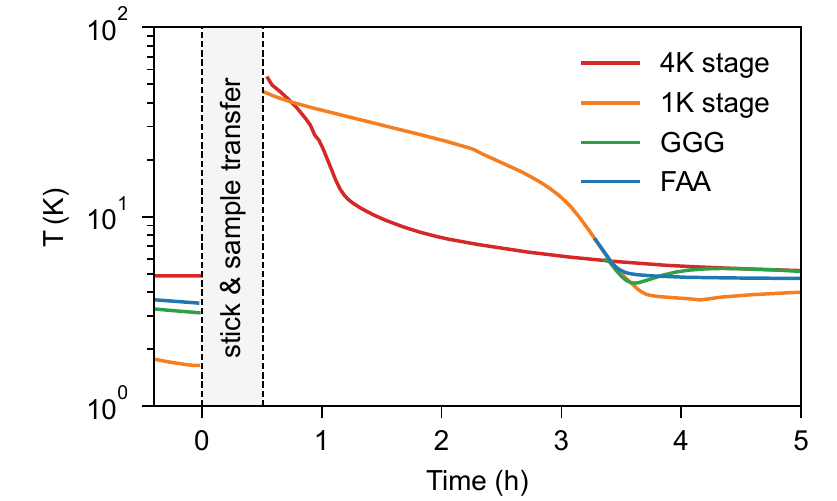}
\caption []{The temperature evolution of the different stages of the mK stick after a sample transfer. During the sample transfer the horizontal sample manipulator is cooled by LN2.}
\label{fig: mK-stick-loading}
\end{figure}

After extracting or loading the sample, the mK stick is loaded back by the reversed manipulation sequence. Fig.~\ref{fig: mK-stick-loading} shows that the full manipulation cycle, i.e. lifting the mK stick, extracting or loading the sample, and loading the mK stick back into the cryostat takes about 30 to 40 minutes, during which the 4K and 1K stages of the mK stick warm up to 40--50 K. A subsequent cooldown of both stages back to 4K takes about 4--5 hours. Note that during the sample manipulation, the mK stick's temperature remains low, which prevents it from outgassing. 

The extraction of the mK stick out of UHV starts with lifting the mK stick up from the cryostat, i.e. executing the steps shown in Fig.~\ref{fig: mK-stick-manipulation}. After reaching the stage shown in Fig.~\ref{fig: mK-stick-manipulation}d the gate valve to the CC is closed, the TC is disconnected from the CC and vented. Then, the frame supporting the UHV system (PC1, PC2 and TC) is moved relative to the CC in the direction perpendicular to the plane of Fig.~\ref{fig: Layout} by a distance of ~900 mm (see the inset of Fig.~\ref{fig: Layout}). After this translation, the space below the TC is free and the manipulator is brought back to the state shown in Fig.~\ref{fig: mK-stick-manipulation}b, so that the mK stick can be removed manually from it as the inset of Fig.~\ref{fig: Layout} suggests.

\subsection{ADR cycle} 
\label{section: ADR cycle}

The ADR technique is well-established, and its detailed description can be found in numerous literature sources \cite{Pobell2007}. Here we give a brief account of it, noting that the cooling is possible due to the magnetocaloric effect in certain paramagnetic materials that possess a large magnetic entropy at low temperatures. Assuming FAA to be a collection of noninteracting magnetic dipoles, one obtains the dependence of its magnetic entropy on the temperature $T$ and the external field $B$ \cite{Wikus2011} as:

\begin{equation}
S(B,T)=nR\left \{x(\coth(x)-y\coth(xy)) +\ln\left(\frac{\sinh(xy)}{\sinh(x)}\right) \right\}
\label{eq: s}
\end{equation}

\noindent with $x=\frac{\mu_\mathrm{B}gB}{2k_\mathrm{B}T}$ and $y=2J+1$. Here $n$ is the number of moles of FAA in the ADR pill, R is the ideal gas constant, $\mu_\mathrm{B}$ is the Bohr magneton, $g$ is the g-factor of an electron, $k_\mathrm{B}$ is the Boltzmann constant, and $J$=5/2 is the total angular momentum of the paramagnetic ions in FAA \cite{Wikus2011}.

\begin{figure}
\centering
\includegraphics [angle=0, width=7.9 cm]{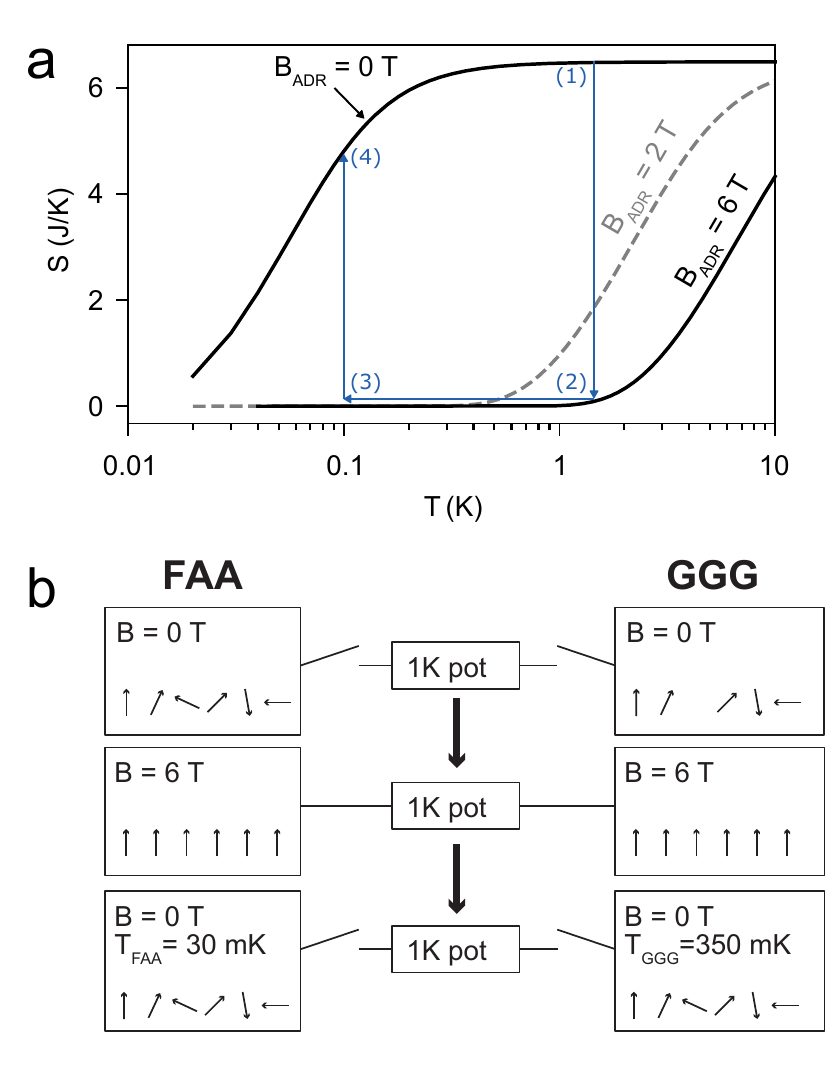}
\caption []{a) $S(B,T)$ of an FAA crystal with mass of 210 g calculated using the non-interacting dipole model \cite{Wikus2011}. The trace (1)-(2)-(3)-(4) indicates a possible ADR cooling cycle in which the isothermal magnetization (1)-(2) is performed at 1.5 K, the adiabatic demagnetization (2)-(3) is stopped at 0.1 K, after which the remaining entropy is used to keep the temperature constant (3)-(4). b) Technical scheme of the two-stage ADR pill employed in our cryostat.}
\label{fig: ADR-basics}
\end{figure}

Fig.~\ref{fig: ADR-basics}a shows plots of $S(B, T)$ calculated with Eq.\ref{eq: s}. The ADR cooling cycle starts at a fixed temperature and zero $B$ field (point 1). In the first step, the $B$ field is increased to a maximum (6 T in our case) while keeping the paramagnetic material in isothermal contact with its environment (point 2). The increase of the field leads to a drop in the magnetic entropy of the paramagnetic material. At the same time, its thermal entropy initially rises, because the heat corresponding to the magnetic entropy before magnetization remains present. However, since the paramagnetic material is in contact with the thermal reservoir of the environment, this heat is transported away (isothermal conditions). In the second step, the paramagnetic material and that part of the environment that is to be refrigerated are decoupled from the thermal reservoir. In this adiabatic condition, the $B$ field is then decreased to zero. While the total entropy remains constant (adiabatic condition), there is a transfer from thermal to magnetic entropy as the magnetic moments overcome the alignment in the $B$ field. This heat transfer leads to a drop in temperature. The ADR base temperature that can be achieved in this way is defined by two parameters: the starting temperature in point 1 and the maximum field value in point 2. Note that the magnetic ordering temperature of FAA defines the absolute lower boundary of temperatures that are attainable by ADR with FAA. Of course, also the heat capacity of the load to be cooled has an effect on the effective base temperature. However, this is usually much smaller than the heat capacity of the system of paramagnetic moments, because the specific heat of FAA in the relevant range of temperatures is much larger than that of e.g. copper \cite{Vilches1966,Pobell2007}. 

Alternatively, stopping the demagnetization at a target temperature before the $B$ field reaches zero (point 3) allows one to hold this temperature constant by slowly decreasing the $B$ field, such that the increase in the magnetic entropy of FAA balances the residual heat flux from the thermal reservoir into the decoupled system (FAA plus load to be cooled). Clearly, the temperature regulation becomes impossible when the $B$ filed reaches zero. The holding time depends on the magnetic entropy of the FAA at the given temperature and the heat leak $\dot{Q}$ (for more details, see section \ref{section: Temperature regulation}).

To establish adiabatic conditions in the mK stick, we employ two commercially acquired UHV-compatible gas-gap heat switches (Fig.~\ref{fig: mK-stick}). Here we show the procedure schematically in Fig. 9b. A gas-gap heat switch typically comprises a thin-walled stainless steel tube containing $^3$He at a pressure of a few mbar and a miniature charcoal pump attached to it. The principle of its operation has been described in detail in the literature \cite{Torre1984}. Briefly, heating the charcoal in the switches to about 15--20 K increases their thermal conductivity to ~10--50 mW/K, thereby establishing a thermal connection between the 1K pot and both stages (FAA and GGG) of the ADR pill. With the heat switches in this ON state, the ADR magnet is ramped up to 6 T. The 1K pot bath removes the heat generated during the magnetization (magnetic entropy is transferred in thermal entropy). When $T_\mathrm{FAA}$ and $T_\mathrm{GGG}$ approach the 1K pot temperature, the heat switches are turned to their OFF state by deactivating the heating of the charcoal. The $^3$He pressure inside the switches decreases, reducing the heat conductivity to $\sim$1~$\mu$W/K (per switch). When the FAA and GGG stages decouple from the 1K pot bath, the final step of demagnetization may start.

\begin{figure}
\centering
\includegraphics [angle=0, width=7.9 cm]{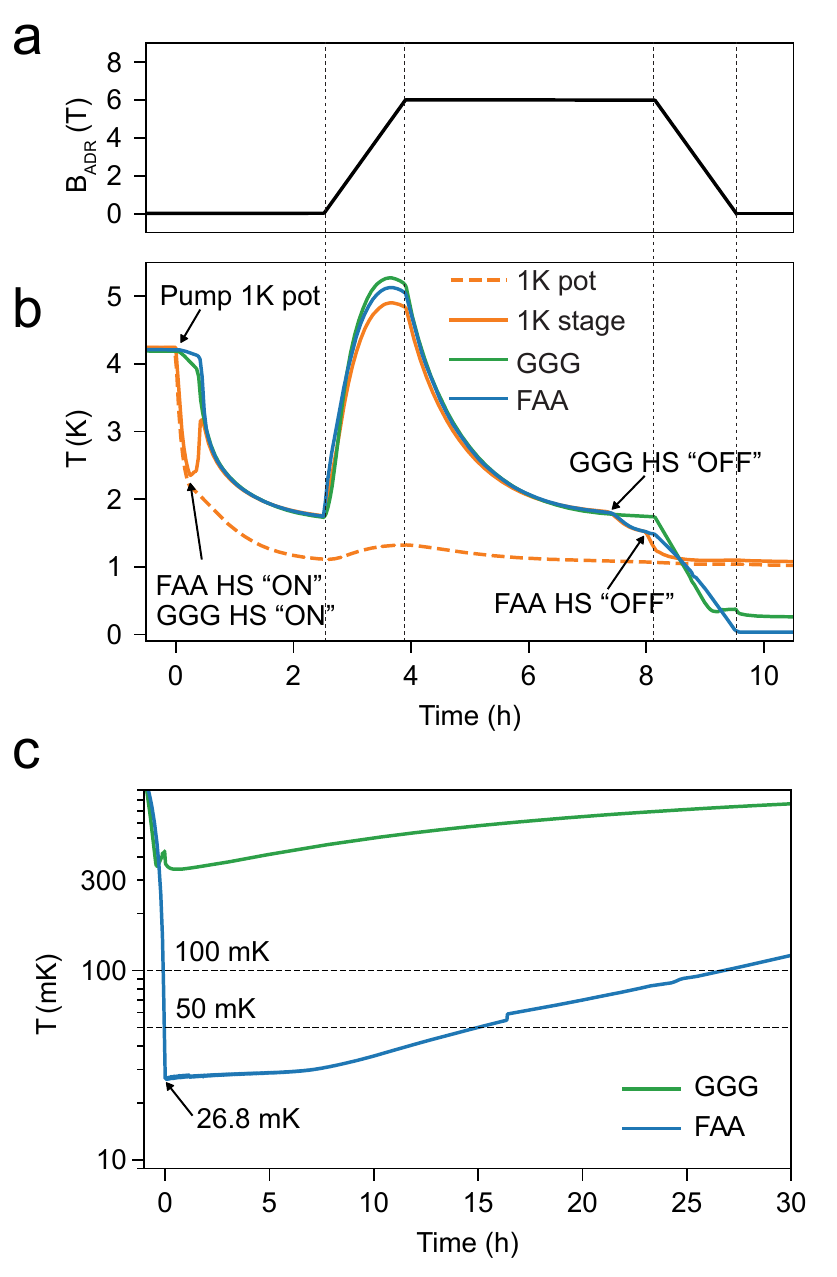}
\caption []{a-b) Typical cooling cycle starting from 4.2 K and finishing at the lowest attainable millikelvin temperature. a) Changes of the $B$ field of the ADR magnet during the ADR cycle and corresponding to the temperature profiles displayed in (b). Changes in $B$ influence the temperature: During magnetization the temperatures increase. After reaching the maximum value, $B$ is kept constant to allow for thermalization of the ADR pill. Demagnetization starts when $T_{\mathrm{FAA}}=1.52$ K and $T_{\mathrm{GGG}}=1.78$ K. The lowest temperature is reached when $B_{\mathrm{ADR}}=0$ after which $T_{\mathrm{FAA}}$ and $T_{\mathrm{GGG}}$ start increasing. c) Evolution of the $T_{\mathrm{FAA}}$ and $T_{\mathrm{GGG}}$ after the end of the ADR cycle shown in (a-b). }
\label{fig: ADR-cycle}
\end{figure}

Fig.~\ref{fig: ADR-cycle} shows an actual ADR run that starts with turning on the pumping of the 1K pot. As becomes apparent from Fig.~\ref{fig: ADR-cycle}b, the whole system needs about 2 h to reach the base temperature near 1 K (see \ref{tab: T}). The following ramp up of the ADR magnet proceeds with the rate of 8 mT/s in order to keep the temperature of the FAA and GGG stages below 5 K. After reaching the maximum field, it takes about 3.5 h until the temperature of the ADR pill drops again to 1.6 K. Next, the GGG heat switch is turned off. Turning off the FAA heat switch with a delay of 0.5 h lets us decrease the FAA temperature further to 1.4 K, because the power dissipated by the heaters of the charcoal in the heat switches only half as large when only one switch remains in the ON state (see Fig.~\ref{fig: ADR-cycle}). With both heat switches in the OFF state, the ADR magnet is demagnetized with the rate of 8 mT/s. This relatively slow demagnetization prevents the generation of excessive heat by eddy currents. However, our experience shows that the demagnetization rate can be increased to at least 20--30 mT/s without noticeable deterioration of the thermal performance.

The demagnetization results in a decrease of the temperature of the ADR pill. The GGG temperature reaches 350 mK when the field of the ADR magnet is about 1~T. At these conditions, GGG experiences an antiferromagnetic ordering transition \cite{Schiffer1994} which precludes further cooling. In contrast, the FAA temperature keeps dropping further and reaches a minimum of 26.8 mK, close to which temperature the electron spins inside FAA also order \cite{Vilches1966}. This point marks the end of the ADR cycle. After reaching its lowest temperature, the ADR setup starts warming up, because of the residual heat leak. Fig.~\ref{fig: ADR-cycle}c shows that it takes $T_\mathrm{FAA}$ about 15 h to reach 50 mK and more than 25 hours to 100 mK. After the warming up, the ADR pill needs to be regenerated by repeating the cycle. In our setup, the described ADR cooling cycle is fully automated and performed without any human intervention.

\subsection{Temperature regulation} 
\label{section: Temperature regulation}

\begin{figure}
\centering
\includegraphics [angle=0, width=7.9 cm]{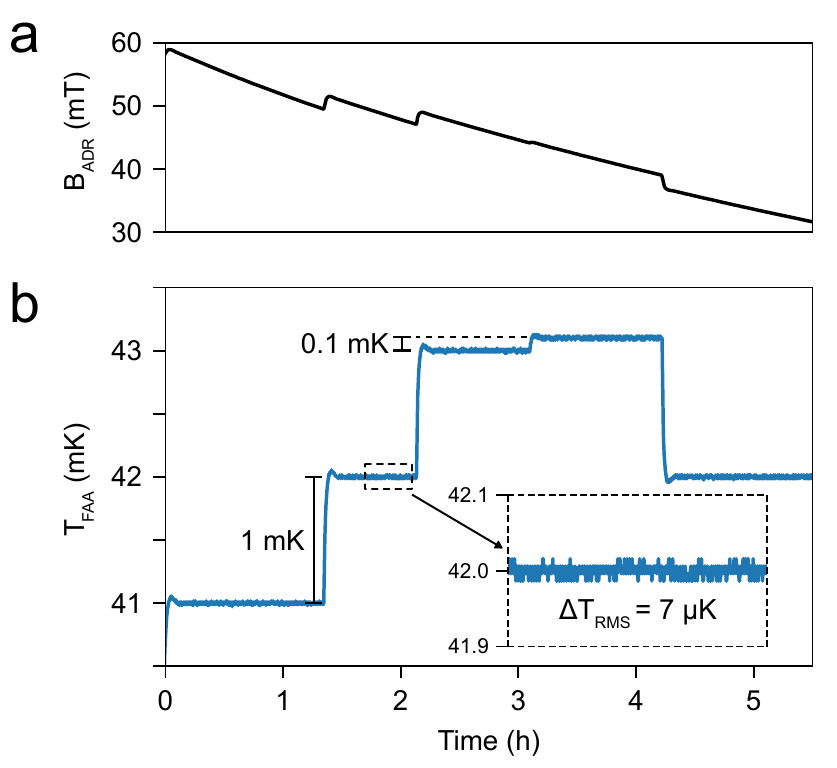}
\caption []{Temperature regulation with ADR. (a) The $B$ field of the ADR magnet as set by the software feedback loop to execute the temperature trace shown in (b). (b) An exemplary trace of the $T_\mathrm{FAA}$ realized by the ADR regulation. Inset shows the magnification of a constant temperature segment to visualize the temperature regulation accuracy.}
\label{fig: ADR-reg}
\end{figure}

\begin{figure}
\centering
\includegraphics [angle=0, width=7.9 cm]{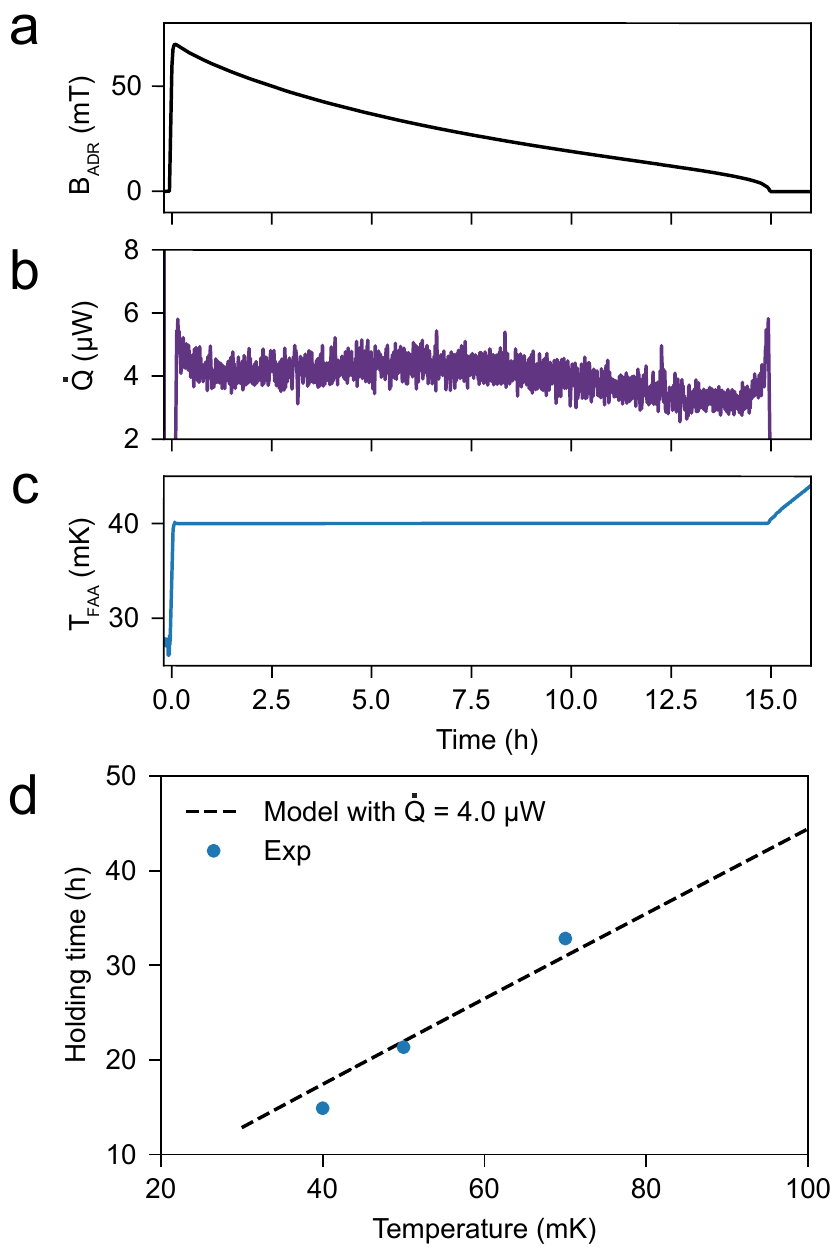}
\caption []{(a) Evolution of the $B$ field produced in the ADR magnet by the software feedback loop for holding $T_\mathrm{FAA}$=40~mK. The regulation starts at the end of the ADR cooling cycle, i.e. after the ADR magnet field has reached zero and the FAA temperature has reached 27~mK. To increase the temperature to 40~mK, a $B$ field of 70~mT is initially applied. The regulation ends when the ADR $B$ field reaches zero again. Note that the regulation at $T>$40~mK at this point remains still possible. (b) The heat leak to the FAA stage at $T=40$~mK, extracted using the data in (a). (c) $T_\mathrm{FAA}$ resulting from the regulation shown in (a). (d) Holding times measured at specific temperatures. Each experimental point was measured by initiating the temperature regulation immediately after the ADR cooling cycle, similar to (a). The dashed line is calculated with the assumption of a heat leak to the FAA stage of 4 $\mu$W. For details, see text.}
\label{fig: ADR-heatleak}
\end{figure}

One of the unique features of ADR is the possibility to precisely control the temperature via regulating the $B$ field. We establish such a control by introducing a software feedback loop that receives an input from the FAA temperature sensor and responds by regulating the current that flows through the ADR magnet's coil. As Fig.~\ref{fig: ADR-reg} shows, within several minutes, the regulation achieves a 7 $\mu$K accuracy in stabilizing a desired temperature in the lower millikelvin range. Notably, regulation works well also at higher temperatures, albeit with a somewhat smaller accuracy.

Using the temperature regulation technique, we also evaluate the residual heat leak towards the FAA stage from the environment. Setting the temperature to 40~mK immediately after finishing the demagnetization cycle, we measure the FAA's holding time at this temperature. As Fig.~\ref{fig: ADR-heatleak} shows, the holding time equals 15~h. Applying the analytical formula for $S(B,T)$ for FAA, we obtain the heat leak $\dot{Q}=TdS/dt \approx 4 \mu$W from the experimental $B(t)$ data. Recalling that the expected heat leak of the FAA heat switch should be about 1 $\mu$W we conclude that the STM and RuOx sensor wiring plus the FAA stage's thermal isolation inside the ADR pill introduce an additional heat leak of 3 $\mu$W.

Calculating the total heat $Q(T_3)$ that can be absorbed by the FAA stage of the ADR pill at the temperature $T_3$ (point 3 in Fig.~\ref{fig: ADR-basics}) as $Q(T_3)=S(0,T_3)-S(B_\mathrm{max},T_3)$ \cite{Wikus2011} and using the obtained value of the heat leak, we predict the holding time of our FAA stage at different temperatures. As Fig.~\ref{fig: ADR-heatleak} shows, the calculated values agree well with the experimentally measured holding time data. The model curve also shows that the holding time at 100 mK may be as long as two days.

\subsection{STM perfomance} 
\label{section: STM perfomance}

\begin{figure}
\centering
\includegraphics [angle=0, width=7.9 cm]{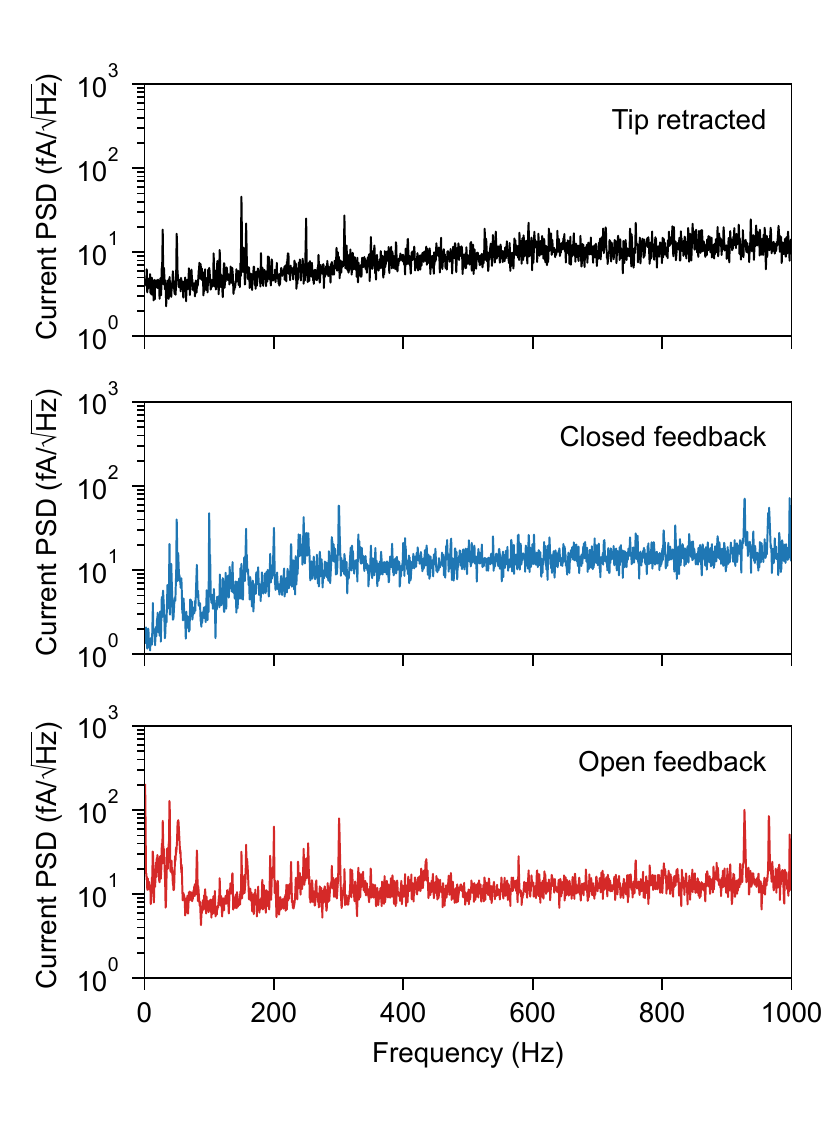}
\caption []{Power spectral density of the tunneling current $I_\mathrm{T}$ acquired at $T_\mathrm{FAA}$ = 29 mK at three different conditions: tip retracted, closed feedback, and open feedback. The data were taken with a PtIr tip on a clean Al(100) surface. The spectra in contact were acquired at a setpoint $I_\mathrm{T}$ = 100 pA and $V$ = 10 mV.}
\label{fig: STM-noise}
\end{figure}

\begin{figure}
\centering
\includegraphics [angle=0, width=7.9 cm]{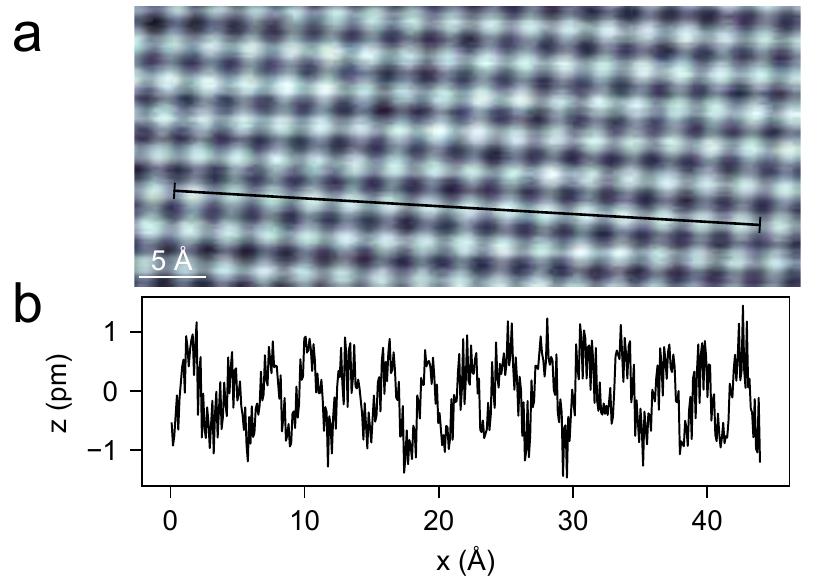}
\caption []{\textit{Top}: Unprocessed constant current STM image of a clean Al(100) surface scanned at $T_\mathrm{FAA}$=196 mK with a clean PtIr tip. The image, exhibiting atomic structure of the surface, has a resolution of 512 pixel/line and was scanned with a speed of 8.1 nm/s. The tunneling setpoint was $I_\mathrm{T}$ = 2 nA and $V$ = 1 mV. \textit{Bottom}: A profile along the black line shown in the image.}
\label{fig: STM-imaging}
\end{figure}

We demonstrate the STM performance by first showing in Fig.~\ref{fig: STM-noise} the power spectral density of the tunneling current measured at 29 mK over an atomically clean Al(100) surface prepared in UHV by repeated cycles of Ar$^{+}$ sputtering and annealing at 500\textdegree{}C. The tunneling current was measured using a fixed $10^9$ gain amplifier from NF Corporation \cite{NF} at a setpoint of $I_\mathrm{T}$ = 100 pA, $V$ = 10 mV. The noise data, as well as the STM images and $dI/dV$ spectra, were acquired with Nanonis SPM control electronics \cite{Nanonis}. The spectra in Fig.~\ref{fig: STM-noise} show that our system reaches a remarkable degree of mechanical stability. The high stability of our STM junction also results in topographic noise smaller than 0.5 pm (peak to peak), as the scanned image of the Al(100) surface demonstrates (Fig.~\ref{fig: STM-imaging}). Finally, neither the warming up of the STM after finishing an ADR cooling cycle nor the ADR regulation of the temperature affect the STM noise.

\subsection{STM junction effective temperature} 
\label{section: STM junction effective temperature}

\begin{figure}
\centering
\includegraphics [angle=0, width=7.9 cm]{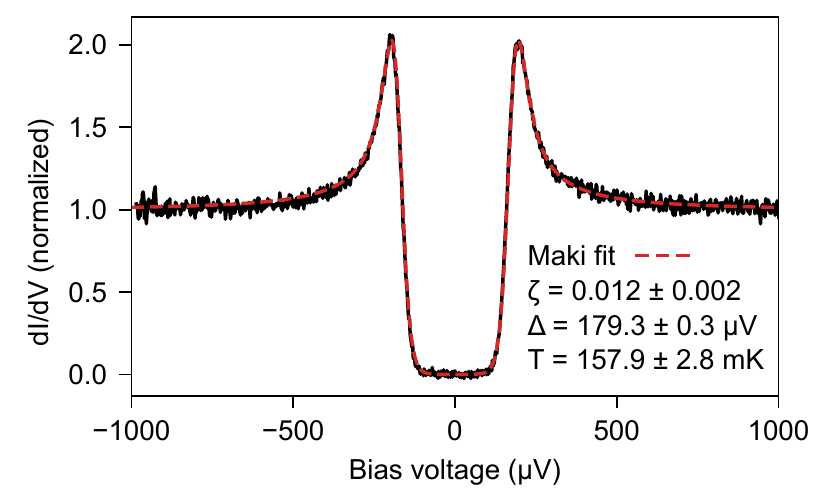}
\caption []{Measurement of the differential tunneling conductance $dI/dV$ on Al(100) at $T_\mathrm{FAA}$ = 27.5 mK using a clean PtIr tip (black curve). The single $dI/dV$ spectrum (raw data) is measured by the lock-in technique with an AC modulation amplitude of 4~$\mu$V and a frequency of 187.7 Hz using a 1/100 bias voltage divider. The acquisition time was 189 s (1024 data points with 150 ms integration time). The tip was stabilized at $I_\mathrm{T}$ = 500 pA and $V$ = 1 mV. The red dashed curve shows the fit based on the Maki function \cite{Assig2013,Machida2018}.}
\label{fig: STS-gap}
\end{figure}

It has become a common practice to demonstrate the effective electronic temperature of mK STM junction by measuring the density of states around a superconducting gap. Such a measurement is necessary because the phonon bath's temperature, measured in our case by the RuOx sensor, may not reflect the electronic temperature. Therefore, we present a scanning tunneling spectroscopy (STS) measurement of the superconducting gap of the Al(100) surface. Fig.~\ref{fig: STS-gap} displays a single $dI/dV$ spectrum (raw data) acquired with the help of the internal lock-in of the Nanonis controller electronics \cite{Nanonis}. We filtered the bias and all five high-voltage lines to the STM piezo to minimize the electric noise. The bias line was filtered with a commercial 5000 pF pi-filter \cite{API_bias} mounted in line with a 3 k$\Omega$ resistor, while for the high-voltage lines we used 4500 pF pi-filters \cite{Filter_piezo}. As one can see in Fig.~\ref{fig: STS-gap}, the Maki fit \cite{Assig2013,Machida2018} of the spectrum comprising the gap and the coherence peaks yields the effective temperature $T_\mathrm{eff}$=157 mK. Although the obtained value is higher than the (phonon bath) temperature of 27.5 mK at which the measurement has been performed, it is within the range of the values reported for the DR-based mK STMs \cite{Song2010,Singh2013,Assig2013,Misra2013,Roychowdhury2014,vonAllworden2018,Machida2018,Balashov2018,Wong2020}.

Surprisingly, however, we find that the electric noise filtering scheme does not substantially affect $T_\mathrm{eff}$ in our case. Removing the bias line filter increases $T_\mathrm{eff}$ to 225 mK. Grounding the temperature sensor wiring does not affect $T_\mathrm{eff}$ at all. At the same time, in agreement with ref. \cite{Schwenk2020}, we see that the type of the current amplifier used for STS influences $T_\mathrm{eff}$: Switching to a Femto DLPCA 200 \cite{Femto} increased $T_\mathrm{eff}$ to 200 mK. 

\section{Conclusion and outlook} 
\label{section: Conclusion and outlook}

In conclusion, we have built -- to the best of our knowledge -- the first UHV STM that uses adiabatic demagnetization of electronic spins to perform measurements at well-controlled millikelvin temperatures. Due to its all-solid-state design, our UHV ADR mK STM is relatively simple, compact and very modular. In particular, the mK stick, which carries the STM and the paramagnetic ADR pill, can be easily extracted from UHV without warming up the main LHe bath cryostat. We expect that this feature of our setup should make its further development and service simple, allowing for an efficient everyday operation. Although the current version of the mK stick is not bakeable due to the low thermal stability of the paramagnetic salt in the ADR-pill, we show that this complication does not preclude work on atomically clean surfaces prepared in UHV. At the same time, the continuing discovery of new magnetocaloric materials makes the prospects of fully bakeable ADR setups reaching 100 mK very realistic \cite{Jang2015}. 

Another technical novelty demonstrated in this work is the successful use of the high-capacity cryopump as a substitution for mechanical pumping of the 1K pot. We find that the cryopump's silent operation mode is essential for reaching the remarkable noise figures demonstrated in our measurements. Finally, we showed that the lowest attainable effective electronic temperature of our STM junction is $T_\mathrm{eff}$ = 157 mK. This value is comparable with the data reported by other groups that use mK STMs based on dilution refrigerators. In our case, $T_\mathrm{eff}$ reacts weakly on the removal of high-frequency noise filters. The factors limiting the $T_\mathrm{eff}$ in our junction will be analyzed in upcoming publications.

\begin{acknowledgments}
We acknowledge fruitful discussions with Josef Baumgartner (Entropy GmbH, Munich, Germany), Simon Chase (Chase Research Cryogenics Ltd, Sheffield, UK), Kurt Haselwimmer, Michael Krzyzowski (CryoVac GmbH \& Co KG, Troisdorf, Germany), George Lecomte (GVL Cryoengineering GmbH, Stolberg, Germany), Gilbert G. Lonzarich (Cavendish Laboratory, University of Cambridge, UK), Jose Martinez Castro (PGI-3, Forschungszentrum J\"ulich), Stefan Stahl (Stahl-Electronics, Mettenheim, Germany), and Doreen Wernicke (Entropy GmbH, Munich, Germany). We also thank Werner H\"urttlen, Jens Prigge, and Helmut Stollwerk for technical support. RT acknowledges the support from the Young Investigator Group program of the Helmholtz Association. 
\end{acknowledgments}

\section*{data availability}
	
The data that support the findings of this study are available from the corresponding author upon reasonable request.

\bibliography{ref}

\begin{thebibliography}{60}%
\makeatletter
\providecommand \@ifxundefined [1]{%
 \@ifx{#1\undefined}
}%
\providecommand \@ifnum [1]{%
 \ifnum #1\expandafter \@firstoftwo
 \else \expandafter \@secondoftwo
 \fi
}%
\providecommand \@ifx [1]{%
 \ifx #1\expandafter \@firstoftwo
 \else \expandafter \@secondoftwo
 \fi
}%
\providecommand \natexlab [1]{#1}%
\providecommand \enquote  [1]{``#1''}%
\providecommand \bibnamefont  [1]{#1}%
\providecommand \bibfnamefont [1]{#1}%
\providecommand \citenamefont [1]{#1}%
\providecommand \href@noop [0]{\@secondoftwo}%
\providecommand \href [0]{\begingroup \@sanitize@url \@href}%
\providecommand \@href[1]{\@@startlink{#1}\@@href}%
\providecommand \@@href[1]{\endgroup#1\@@endlink}%
\providecommand \@sanitize@url [0]{\catcode `\\12\catcode `\$12\catcode
  `\&12\catcode `\#12\catcode `\^12\catcode `\_12\catcode `\%12\relax}%
\providecommand \@@startlink[1]{}%
\providecommand \@@endlink[0]{}%
\providecommand \url  [0]{\begingroup\@sanitize@url \@url }%
\providecommand \@url [1]{\endgroup\@href {#1}{\urlprefix }}%
\providecommand \urlprefix  [0]{URL }%
\providecommand \Eprint [0]{\href }%
\providecommand \doibase [0]{http://dx.doi.org/}%
\providecommand \selectlanguage [0]{\@gobble}%
\providecommand \bibinfo  [0]{\@secondoftwo}%
\providecommand \bibfield  [0]{\@secondoftwo}%
\providecommand \translation [1]{[#1]}%
\providecommand \BibitemOpen [0]{}%
\providecommand \bibitemStop [0]{}%
\providecommand \bibitemNoStop [0]{.\EOS\space}%
\providecommand \EOS [0]{\spacefactor3000\relax}%
\providecommand \BibitemShut  [1]{\csname bibitem#1\endcsname}%
\let\auto@bib@innerbib\@empty
\bibitem [{\citenamefont {Niimi}, \citenamefont {Kambara},\ and\ \citenamefont
  {Fukuyama}(2009)}]{Niimi2009}%
  \BibitemOpen
  \bibfield  {author} {\bibinfo {author} {\bibfnamefont {Y.}~\bibnamefont
  {Niimi}}, \bibinfo {author} {\bibfnamefont {H.}~\bibnamefont {Kambara}}, \
  and\ \bibinfo {author} {\bibfnamefont {H.}~\bibnamefont {Fukuyama}},\
  }\bibfield  {title} {\enquote {\bibinfo {title} {Localized distributions of
  quasi-two-dimensional electronic states near defects artificially created at
  graphite surfaces in magnetic fields},}\ }\href {\doibase
  10.1103/PhysRevLett.102.026803} {\bibfield  {journal} {\bibinfo  {journal}
  {Phys. Rev. Lett.}\ }\textbf {\bibinfo {volume} {102}},\ \bibinfo {pages}
  {026803} (\bibinfo {year} {2009})}\BibitemShut {NoStop}%
\bibitem [{\citenamefont {Song}\ \emph
  {et~al.}(2010{\natexlab{a}})\citenamefont {Song}, \citenamefont {Otte},
  \citenamefont {Kuk}, \citenamefont {Hu}, \citenamefont {Torrance},
  \citenamefont {First}, \citenamefont {de~Heer}, \citenamefont {Min},
  \citenamefont {Adam}, \citenamefont {Stiles}, \citenamefont {MacDonald},\
  and\ \citenamefont {Stroscio}}]{Song2010_1}%
  \BibitemOpen
  \bibfield  {author} {\bibinfo {author} {\bibfnamefont {Y.~J.}\ \bibnamefont
  {Song}}, \bibinfo {author} {\bibfnamefont {A.~F.}\ \bibnamefont {Otte}},
  \bibinfo {author} {\bibfnamefont {Y.}~\bibnamefont {Kuk}}, \bibinfo {author}
  {\bibfnamefont {Y.}~\bibnamefont {Hu}}, \bibinfo {author} {\bibfnamefont
  {D.~B.}\ \bibnamefont {Torrance}}, \bibinfo {author} {\bibfnamefont {P.~N.}\
  \bibnamefont {First}}, \bibinfo {author} {\bibfnamefont {W.~A.}\ \bibnamefont
  {de~Heer}}, \bibinfo {author} {\bibfnamefont {H.}~\bibnamefont {Min}},
  \bibinfo {author} {\bibfnamefont {S.}~\bibnamefont {Adam}}, \bibinfo {author}
  {\bibfnamefont {M.~D.}\ \bibnamefont {Stiles}}, \bibinfo {author}
  {\bibfnamefont {A.~H.}\ \bibnamefont {MacDonald}}, \ and\ \bibinfo {author}
  {\bibfnamefont {J.~A.}\ \bibnamefont {Stroscio}},\ }\bibfield  {title}
  {\enquote {\bibinfo {title} {High-resolution tunnelling spectroscopy of a
  graphene quartet},}\ }\href {\doibase 10.1038/nature09330} {\bibfield
  {journal} {\bibinfo  {journal} {Nature}\ }\textbf {\bibinfo {volume} {467}},\
  \bibinfo {pages} {185--189} (\bibinfo {year}
  {2010}{\natexlab{a}})}\BibitemShut {NoStop}%
\bibitem [{\citenamefont {Levy}\ \emph {et~al.}(2013)\citenamefont {Levy},
  \citenamefont {Zhang}, \citenamefont {Ha}, \citenamefont {Sharifi},
  \citenamefont {Talin}, \citenamefont {Kuk},\ and\ \citenamefont
  {Stroscio}}]{Levy2013}%
  \BibitemOpen
  \bibfield  {author} {\bibinfo {author} {\bibfnamefont {N.}~\bibnamefont
  {Levy}}, \bibinfo {author} {\bibfnamefont {T.}~\bibnamefont {Zhang}},
  \bibinfo {author} {\bibfnamefont {J.}~\bibnamefont {Ha}}, \bibinfo {author}
  {\bibfnamefont {F.}~\bibnamefont {Sharifi}}, \bibinfo {author} {\bibfnamefont
  {A.~A.}\ \bibnamefont {Talin}}, \bibinfo {author} {\bibfnamefont
  {Y.}~\bibnamefont {Kuk}}, \ and\ \bibinfo {author} {\bibfnamefont {J.~A.}\
  \bibnamefont {Stroscio}},\ }\bibfield  {title} {\enquote {\bibinfo {title}
  {Experimental evidence for $s$-wave pairing symmetry in superconducting
  {${\mathrm{Cu}}_{x}{\mathrm{Bi}}_{2}{\mathrm{Se}}_{3}$} single crystals using
  a scanning tunneling microscope},}\ }\href {\doibase
  10.1103/PhysRevLett.110.117001} {\bibfield  {journal} {\bibinfo  {journal}
  {Phys. Rev. Lett.}\ }\textbf {\bibinfo {volume} {110}},\ \bibinfo {pages}
  {117001} (\bibinfo {year} {2013})}\BibitemShut {NoStop}%
\bibitem [{\citenamefont {Eltschka}\ \emph {et~al.}(2014)\citenamefont
  {Eltschka}, \citenamefont {J\"ack}, \citenamefont {Assig}, \citenamefont
  {Kondrashov}, \citenamefont {Skvortsov}, \citenamefont {Etzkorn},
  \citenamefont {Ast},\ and\ \citenamefont {Kern}}]{Eltschka2014}%
  \BibitemOpen
  \bibfield  {author} {\bibinfo {author} {\bibfnamefont {M.}~\bibnamefont
  {Eltschka}}, \bibinfo {author} {\bibfnamefont {B.}~\bibnamefont {J\"ack}},
  \bibinfo {author} {\bibfnamefont {M.}~\bibnamefont {Assig}}, \bibinfo
  {author} {\bibfnamefont {O.~V.}\ \bibnamefont {Kondrashov}}, \bibinfo
  {author} {\bibfnamefont {M.~A.}\ \bibnamefont {Skvortsov}}, \bibinfo {author}
  {\bibfnamefont {M.}~\bibnamefont {Etzkorn}}, \bibinfo {author} {\bibfnamefont
  {C.~R.}\ \bibnamefont {Ast}}, \ and\ \bibinfo {author} {\bibfnamefont
  {K.}~\bibnamefont {Kern}},\ }\bibfield  {title} {\enquote {\bibinfo {title}
  {Probing absolute spin polarization at the nanoscale},}\ }\href {\doibase
  10.1021/nl5037947} {\bibfield  {journal} {\bibinfo  {journal} {Nano Lett.}\
  }\textbf {\bibinfo {volume} {14}},\ \bibinfo {pages} {7171--7174} (\bibinfo
  {year} {2014})}\BibitemShut {NoStop}%
\bibitem [{\citenamefont {Roychowdhury}\ \emph {et~al.}(2015)\citenamefont
  {Roychowdhury}, \citenamefont {Dreyer}, \citenamefont {Anderson},
  \citenamefont {Lobb},\ and\ \citenamefont {Wellstood}}]{Roychowdhury2015}%
  \BibitemOpen
  \bibfield  {author} {\bibinfo {author} {\bibfnamefont {A.}~\bibnamefont
  {Roychowdhury}}, \bibinfo {author} {\bibfnamefont {M.}~\bibnamefont
  {Dreyer}}, \bibinfo {author} {\bibfnamefont {J.~R.}\ \bibnamefont
  {Anderson}}, \bibinfo {author} {\bibfnamefont {C.~J.}\ \bibnamefont {Lobb}},
  \ and\ \bibinfo {author} {\bibfnamefont {F.~C.}\ \bibnamefont {Wellstood}},\
  }\bibfield  {title} {\enquote {\bibinfo {title} {Microwave photon-assisted
  incoherent cooper-pair tunneling in a {Josephson STM}},}\ }\href {\doibase
  10.1103/PhysRevApplied.4.034011} {\bibfield  {journal} {\bibinfo  {journal}
  {Phys. Rev. Applied}\ }\textbf {\bibinfo {volume} {4}},\ \bibinfo {pages}
  {034011} (\bibinfo {year} {2015})}\BibitemShut {NoStop}%
\bibitem [{\citenamefont {Sun}\ \emph {et~al.}(2015)\citenamefont {Sun},
  \citenamefont {Enayat}, \citenamefont {Maldonado}, \citenamefont {Lithgow},
  \citenamefont {Yelland}, \citenamefont {Peets}, \citenamefont {Yaresko},
  \citenamefont {Schnyder},\ and\ \citenamefont {Wahl}}]{Sun2015}%
  \BibitemOpen
  \bibfield  {author} {\bibinfo {author} {\bibfnamefont {Z.}~\bibnamefont
  {Sun}}, \bibinfo {author} {\bibfnamefont {M.}~\bibnamefont {Enayat}},
  \bibinfo {author} {\bibfnamefont {A.}~\bibnamefont {Maldonado}}, \bibinfo
  {author} {\bibfnamefont {C.}~\bibnamefont {Lithgow}}, \bibinfo {author}
  {\bibfnamefont {E.}~\bibnamefont {Yelland}}, \bibinfo {author} {\bibfnamefont
  {D.~C.}\ \bibnamefont {Peets}}, \bibinfo {author} {\bibfnamefont
  {A.}~\bibnamefont {Yaresko}}, \bibinfo {author} {\bibfnamefont {A.~P.}\
  \bibnamefont {Schnyder}}, \ and\ \bibinfo {author} {\bibfnamefont
  {P.}~\bibnamefont {Wahl}},\ }\bibfield  {title} {\enquote {\bibinfo {title}
  {Dirac surface states and nature of superconductivity in noncentrosymmetric
  {BiPd}},}\ }\href {\doibase 10.1038/ncomms7633} {\bibfield  {journal}
  {\bibinfo  {journal} {Nat. Commun.}\ }\textbf {\bibinfo {volume} {6}},\
  \bibinfo {pages} {6633} (\bibinfo {year} {2015})}\BibitemShut {NoStop}%
\bibitem [{\citenamefont {J\"{a}ck}\ \emph {et~al.}(2015)\citenamefont
  {J\"{a}ck}, \citenamefont {Eltschka}, \citenamefont {Assig}, \citenamefont
  {Hardock}, \citenamefont {Etzkorn}, \citenamefont {Ast},\ and\ \citenamefont
  {Kern}}]{Jack2015}%
  \BibitemOpen
  \bibfield  {author} {\bibinfo {author} {\bibfnamefont {B.}~\bibnamefont
  {J\"{a}ck}}, \bibinfo {author} {\bibfnamefont {M.}~\bibnamefont {Eltschka}},
  \bibinfo {author} {\bibfnamefont {M.}~\bibnamefont {Assig}}, \bibinfo
  {author} {\bibfnamefont {A.}~\bibnamefont {Hardock}}, \bibinfo {author}
  {\bibfnamefont {M.}~\bibnamefont {Etzkorn}}, \bibinfo {author} {\bibfnamefont
  {C.~R.}\ \bibnamefont {Ast}}, \ and\ \bibinfo {author} {\bibfnamefont
  {K.}~\bibnamefont {Kern}},\ }\bibfield  {title} {\enquote {\bibinfo {title}
  {A nanoscale gigahertz source realized with {Josephson} scanning tunneling
  microscopy},}\ }\href {\doibase 10.1063/1.4905322} {\bibfield  {journal}
  {\bibinfo  {journal} {Appl. Phys. Lett.}\ }\textbf {\bibinfo {volume}
  {106}},\ \bibinfo {pages} {013109} (\bibinfo {year} {2015})}\BibitemShut
  {NoStop}%
\bibitem [{\citenamefont {Ast}\ \emph {et~al.}(2016)\citenamefont {Ast},
  \citenamefont {J\"ack}, \citenamefont {Senkpiel}, \citenamefont {Eltschka},
  \citenamefont {Etzkorn}, \citenamefont {Ankerhold},\ and\ \citenamefont
  {Kern}}]{Ast2016}%
  \BibitemOpen
  \bibfield  {author} {\bibinfo {author} {\bibfnamefont {C.~R.}\ \bibnamefont
  {Ast}}, \bibinfo {author} {\bibfnamefont {B.}~\bibnamefont {J\"ack}},
  \bibinfo {author} {\bibfnamefont {J.}~\bibnamefont {Senkpiel}}, \bibinfo
  {author} {\bibfnamefont {M.}~\bibnamefont {Eltschka}}, \bibinfo {author}
  {\bibfnamefont {M.}~\bibnamefont {Etzkorn}}, \bibinfo {author} {\bibfnamefont
  {J.}~\bibnamefont {Ankerhold}}, \ and\ \bibinfo {author} {\bibfnamefont
  {K.}~\bibnamefont {Kern}},\ }\bibfield  {title} {\enquote {\bibinfo {title}
  {Sensing the quantum limit in scanning tunnelling spectroscopy},}\ }\href
  {\doibase 10.1038/ncomms13009} {\bibfield  {journal} {\bibinfo  {journal}
  {Nat. Commun.}\ }\textbf {\bibinfo {volume} {7}},\ \bibinfo {pages} {13009}
  (\bibinfo {year} {2016})}\BibitemShut {NoStop}%
\bibitem [{\citenamefont {J\"{a}ck}\ \emph {et~al.}(2016)\citenamefont
  {J\"{a}ck}, \citenamefont {Eltschka}, \citenamefont {Assig}, \citenamefont
  {Etzkorn}, \citenamefont {Ast},\ and\ \citenamefont {Kern}}]{Jack2016}%
  \BibitemOpen
  \bibfield  {author} {\bibinfo {author} {\bibfnamefont {B.}~\bibnamefont
  {J\"{a}ck}}, \bibinfo {author} {\bibfnamefont {M.}~\bibnamefont {Eltschka}},
  \bibinfo {author} {\bibfnamefont {M.}~\bibnamefont {Assig}}, \bibinfo
  {author} {\bibfnamefont {M.}~\bibnamefont {Etzkorn}}, \bibinfo {author}
  {\bibfnamefont {C.~R.}\ \bibnamefont {Ast}}, \ and\ \bibinfo {author}
  {\bibfnamefont {K.}~\bibnamefont {Kern}},\ }\bibfield  {title} {\enquote
  {\bibinfo {title} {Critical {Josephson} current in the dynamical coulomb
  blockade regime},}\ }\href {\doibase 10.1103/PhysRevB.93.020504} {\bibfield
  {journal} {\bibinfo  {journal} {Phys. Rev. B}\ }\textbf {\bibinfo {volume}
  {93}},\ \bibinfo {pages} {020504} (\bibinfo {year} {2016})}\BibitemShut
  {NoStop}%
\bibitem [{\citenamefont {Natterer}\ \emph {et~al.}(2016)\citenamefont
  {Natterer}, \citenamefont {Ha}, \citenamefont {Baek}, \citenamefont {Zhang},
  \citenamefont {Cullen}, \citenamefont {Zhitenev}, \citenamefont {Kuk},\ and\
  \citenamefont {Stroscio}}]{Natterer2016}%
  \BibitemOpen
  \bibfield  {author} {\bibinfo {author} {\bibfnamefont {F.~D.}\ \bibnamefont
  {Natterer}}, \bibinfo {author} {\bibfnamefont {J.}~\bibnamefont {Ha}},
  \bibinfo {author} {\bibfnamefont {H.}~\bibnamefont {Baek}}, \bibinfo {author}
  {\bibfnamefont {D.}~\bibnamefont {Zhang}}, \bibinfo {author} {\bibfnamefont
  {W.~G.}\ \bibnamefont {Cullen}}, \bibinfo {author} {\bibfnamefont {N.~B.}\
  \bibnamefont {Zhitenev}}, \bibinfo {author} {\bibfnamefont {Y.}~\bibnamefont
  {Kuk}}, \ and\ \bibinfo {author} {\bibfnamefont {J.~A.}\ \bibnamefont
  {Stroscio}},\ }\bibfield  {title} {\enquote {\bibinfo {title} {Scanning
  tunneling spectroscopy of proximity superconductivity in epitaxial multilayer
  graphene},}\ }\href {\doibase 10.1103/PhysRevB.93.045406} {\bibfield
  {journal} {\bibinfo  {journal} {Phys. Rev. B}\ }\textbf {\bibinfo {volume}
  {93}},\ \bibinfo {pages} {045406} (\bibinfo {year} {2016})}\BibitemShut
  {NoStop}%
\bibitem [{\citenamefont {Hamidian}\ \emph {et~al.}(2016)\citenamefont
  {Hamidian}, \citenamefont {Edkins}, \citenamefont {Joo}, \citenamefont
  {Kostin}, \citenamefont {Eisaki}, \citenamefont {Uchida}, \citenamefont
  {Lawler}, \citenamefont {Kim}, \citenamefont {Mackenzie}, \citenamefont
  {Fujita}, \citenamefont {Lee},\ and\ \citenamefont {Davis}}]{Hamidian2016}%
  \BibitemOpen
  \bibfield  {author} {\bibinfo {author} {\bibfnamefont {M.~H.}\ \bibnamefont
  {Hamidian}}, \bibinfo {author} {\bibfnamefont {S.~D.}\ \bibnamefont
  {Edkins}}, \bibinfo {author} {\bibfnamefont {S.~H.}\ \bibnamefont {Joo}},
  \bibinfo {author} {\bibfnamefont {A.}~\bibnamefont {Kostin}}, \bibinfo
  {author} {\bibfnamefont {H.}~\bibnamefont {Eisaki}}, \bibinfo {author}
  {\bibfnamefont {S.}~\bibnamefont {Uchida}}, \bibinfo {author} {\bibfnamefont
  {M.~J.}\ \bibnamefont {Lawler}}, \bibinfo {author} {\bibfnamefont {E.-A.}\
  \bibnamefont {Kim}}, \bibinfo {author} {\bibfnamefont {A.~P.}\ \bibnamefont
  {Mackenzie}}, \bibinfo {author} {\bibfnamefont {K.}~\bibnamefont {Fujita}},
  \bibinfo {author} {\bibfnamefont {J.}~\bibnamefont {Lee}}, \ and\ \bibinfo
  {author} {\bibfnamefont {J.~C.~S.}\ \bibnamefont {Davis}},\ }\bibfield
  {title} {\enquote {\bibinfo {title} {Detection of a {Cooper}-pair density
  wave in
  {${\mathrm{Bi}}_{2}{\mathrm{Sr}}_{2}{\mathrm{CaCu}}_{2}{\mathrm{O}}_{8+\ensuremath{x}}$}},}\
  }\href {\doibase 10.1038/nature17411} {\bibfield  {journal} {\bibinfo
  {journal} {Nature}\ }\textbf {\bibinfo {volume} {532}},\ \bibinfo {pages}
  {343--347} (\bibinfo {year} {2016})}\BibitemShut {NoStop}%
\bibitem [{\citenamefont {Feldman}\ \emph {et~al.}(2017)\citenamefont
  {Feldman}, \citenamefont {Randeria}, \citenamefont {Li}, \citenamefont
  {Jeon}, \citenamefont {Xie}, \citenamefont {Wang}, \citenamefont {Drozdov},
  \citenamefont {Andrei~Bernevig},\ and\ \citenamefont
  {Yazdani}}]{Feldman2017}%
  \BibitemOpen
  \bibfield  {author} {\bibinfo {author} {\bibfnamefont {B.~E.}\ \bibnamefont
  {Feldman}}, \bibinfo {author} {\bibfnamefont {M.~T.}\ \bibnamefont
  {Randeria}}, \bibinfo {author} {\bibfnamefont {J.}~\bibnamefont {Li}},
  \bibinfo {author} {\bibfnamefont {S.}~\bibnamefont {Jeon}}, \bibinfo {author}
  {\bibfnamefont {Y.}~\bibnamefont {Xie}}, \bibinfo {author} {\bibfnamefont
  {Z.}~\bibnamefont {Wang}}, \bibinfo {author} {\bibfnamefont {I.~K.}\
  \bibnamefont {Drozdov}}, \bibinfo {author} {\bibfnamefont {B.}~\bibnamefont
  {Andrei~Bernevig}}, \ and\ \bibinfo {author} {\bibfnamefont {A.}~\bibnamefont
  {Yazdani}},\ }\bibfield  {title} {\enquote {\bibinfo {title} {High-resolution
  studies of the {Majorana} atomic chain platform},}\ }\href {\doibase
  10.1038/nphys3947} {\bibfield  {journal} {\bibinfo  {journal} {Nat. Phys.}\
  }\textbf {\bibinfo {volume} {13}},\ \bibinfo {pages} {286--291} (\bibinfo
  {year} {2017})}\BibitemShut {NoStop}%
\bibitem [{\citenamefont {Clark}\ \emph {et~al.}(2018)\citenamefont {Clark},
  \citenamefont {Neat}, \citenamefont {Okawa}, \citenamefont {Bawden},
  \citenamefont {Markovi{\'c}}, \citenamefont {Mazzola}, \citenamefont {Feng},
  \citenamefont {Sunko}, \citenamefont {Riley}, \citenamefont {Meevasana},
  \citenamefont {Fujii}, \citenamefont {Vobornik}, \citenamefont {Kim},
  \citenamefont {Hoesch}, \citenamefont {Sasagawa}, \citenamefont {Wahl},
  \citenamefont {Bahramy},\ and\ \citenamefont {King}}]{Clark2018}%
  \BibitemOpen
  \bibfield  {author} {\bibinfo {author} {\bibfnamefont {O.~J.}\ \bibnamefont
  {Clark}}, \bibinfo {author} {\bibfnamefont {M.~J.}\ \bibnamefont {Neat}},
  \bibinfo {author} {\bibfnamefont {K.}~\bibnamefont {Okawa}}, \bibinfo
  {author} {\bibfnamefont {L.}~\bibnamefont {Bawden}}, \bibinfo {author}
  {\bibfnamefont {I.}~\bibnamefont {Markovi{\'c}}}, \bibinfo {author}
  {\bibfnamefont {F.}~\bibnamefont {Mazzola}}, \bibinfo {author} {\bibfnamefont
  {J.}~\bibnamefont {Feng}}, \bibinfo {author} {\bibfnamefont {V.}~\bibnamefont
  {Sunko}}, \bibinfo {author} {\bibfnamefont {J.~M.}\ \bibnamefont {Riley}},
  \bibinfo {author} {\bibfnamefont {W.}~\bibnamefont {Meevasana}}, \bibinfo
  {author} {\bibfnamefont {J.}~\bibnamefont {Fujii}}, \bibinfo {author}
  {\bibfnamefont {I.}~\bibnamefont {Vobornik}}, \bibinfo {author}
  {\bibfnamefont {T.~K.}\ \bibnamefont {Kim}}, \bibinfo {author} {\bibfnamefont
  {M.}~\bibnamefont {Hoesch}}, \bibinfo {author} {\bibfnamefont
  {T.}~\bibnamefont {Sasagawa}}, \bibinfo {author} {\bibfnamefont
  {P.}~\bibnamefont {Wahl}}, \bibinfo {author} {\bibfnamefont {M.~S.}\
  \bibnamefont {Bahramy}}, \ and\ \bibinfo {author} {\bibfnamefont {P.~D.~C.}\
  \bibnamefont {King}},\ }\bibfield  {title} {\enquote {\bibinfo {title}
  {Fermiology and superconductivity of topological surface states in
  {${\mathrm{PdTe}}_{2}$}},}\ }\href {\doibase 10.1103/PhysRevLett.120.156401}
  {\bibfield  {journal} {\bibinfo  {journal} {Phys. Rev. Lett.}\ }\textbf
  {\bibinfo {volume} {120}},\ \bibinfo {pages} {156401} (\bibinfo {year}
  {2018})}\BibitemShut {NoStop}%
\bibitem [{\citenamefont {Machida}\ \emph {et~al.}(2019)\citenamefont
  {Machida}, \citenamefont {Sun}, \citenamefont {Pyon}, \citenamefont {Takeda},
  \citenamefont {Kohsaka}, \citenamefont {Hanaguri}, \citenamefont {Sasagawa},\
  and\ \citenamefont {Tamegai}}]{Machida2019}%
  \BibitemOpen
  \bibfield  {author} {\bibinfo {author} {\bibfnamefont {T.}~\bibnamefont
  {Machida}}, \bibinfo {author} {\bibfnamefont {Y.}~\bibnamefont {Sun}},
  \bibinfo {author} {\bibfnamefont {S.}~\bibnamefont {Pyon}}, \bibinfo {author}
  {\bibfnamefont {S.}~\bibnamefont {Takeda}}, \bibinfo {author} {\bibfnamefont
  {Y.}~\bibnamefont {Kohsaka}}, \bibinfo {author} {\bibfnamefont
  {T.}~\bibnamefont {Hanaguri}}, \bibinfo {author} {\bibfnamefont
  {T.}~\bibnamefont {Sasagawa}}, \ and\ \bibinfo {author} {\bibfnamefont
  {T.}~\bibnamefont {Tamegai}},\ }\bibfield  {title} {\enquote {\bibinfo
  {title} {Zero-energy vortex bound state in the superconducting topological
  surface state of {Fe(Se,Te)}},}\ }\href {\doibase 10.1038/s41563-019-0397-1}
  {\bibfield  {journal} {\bibinfo  {journal} {Nat. Mater.}\ }\textbf {\bibinfo
  {volume} {18}},\ \bibinfo {pages} {811--815} (\bibinfo {year}
  {2019})}\BibitemShut {NoStop}%
\bibitem [{\citenamefont {Senkpiel}\ \emph {et~al.}(2020)\citenamefont
  {Senkpiel}, \citenamefont {Kl{\"o}ckner}, \citenamefont {Etzkorn},
  \citenamefont {Dambach}, \citenamefont {Kubala}, \citenamefont {Belzig},
  \citenamefont {Yeyati}, \citenamefont {Cuevas}, \citenamefont {Pauly},
  \citenamefont {Ankerhold}, \citenamefont {Ast},\ and\ \citenamefont
  {Kern}}]{Senkpiel2020}%
  \BibitemOpen
  \bibfield  {author} {\bibinfo {author} {\bibfnamefont {J.}~\bibnamefont
  {Senkpiel}}, \bibinfo {author} {\bibfnamefont {J.~C.}\ \bibnamefont
  {Kl{\"o}ckner}}, \bibinfo {author} {\bibfnamefont {M.}~\bibnamefont
  {Etzkorn}}, \bibinfo {author} {\bibfnamefont {S.}~\bibnamefont {Dambach}},
  \bibinfo {author} {\bibfnamefont {B.}~\bibnamefont {Kubala}}, \bibinfo
  {author} {\bibfnamefont {W.}~\bibnamefont {Belzig}}, \bibinfo {author}
  {\bibfnamefont {A.~L.}\ \bibnamefont {Yeyati}}, \bibinfo {author}
  {\bibfnamefont {J.~C.}\ \bibnamefont {Cuevas}}, \bibinfo {author}
  {\bibfnamefont {F.}~\bibnamefont {Pauly}}, \bibinfo {author} {\bibfnamefont
  {J.}~\bibnamefont {Ankerhold}}, \bibinfo {author} {\bibfnamefont {C.~R.}\
  \bibnamefont {Ast}}, \ and\ \bibinfo {author} {\bibfnamefont
  {K.}~\bibnamefont {Kern}},\ }\bibfield  {title} {\enquote {\bibinfo {title}
  {Dynamical coulomb blockade as a local probe for quantum transport},}\ }\href
  {\doibase 10.1103/PhysRevLett.124.156803} {\bibfield  {journal} {\bibinfo
  {journal} {Phys. Rev. Lett.}\ }\textbf {\bibinfo {volume} {124}},\ \bibinfo
  {pages} {156803} (\bibinfo {year} {2020})}\BibitemShut {NoStop}%
\bibitem [{\citenamefont {van Weerdenburg}\ \emph {et~al.}(2020)\citenamefont
  {van Weerdenburg}, \citenamefont {Steinbrecher}, \citenamefont {van
  Mullekom}, \citenamefont {Gerritsen}, \citenamefont {von Allw\"orden},
  \citenamefont {Natterer},\ and\ \citenamefont
  {Khajetoorians}}]{Weerdenburg2020}%
  \BibitemOpen
  \bibfield  {author} {\bibinfo {author} {\bibfnamefont {W.~M.~J.}\
  \bibnamefont {van Weerdenburg}}, \bibinfo {author} {\bibfnamefont
  {M.}~\bibnamefont {Steinbrecher}}, \bibinfo {author} {\bibfnamefont
  {N.~P.~E.}\ \bibnamefont {van Mullekom}}, \bibinfo {author} {\bibfnamefont
  {J.~W.}\ \bibnamefont {Gerritsen}}, \bibinfo {author} {\bibfnamefont
  {H.}~\bibnamefont {von Allw\"orden}}, \bibinfo {author} {\bibfnamefont
  {F.~D.}\ \bibnamefont {Natterer}}, \ and\ \bibinfo {author} {\bibfnamefont
  {A.~A.}\ \bibnamefont {Khajetoorians}},\ }\href@noop {} {\enquote {\bibinfo
  {title} {A scanning tunneling microscope capable of electron spin resonance
  and pump-probe spectroscopy at {mK} temperature and in vector magnetic
  field},}\ } (\bibinfo {year} {2020}),\ \Eprint
  {http://arxiv.org/abs/2007.01835} {arXiv:2007.01835 [cond-mat.mes-hall]}
  \BibitemShut {NoStop}%
\bibitem [{\citenamefont {Nuckolls}\ \emph {et~al.}(2020)\citenamefont
  {Nuckolls}, \citenamefont {Oh}, \citenamefont {Wong}, \citenamefont {Lian},
  \citenamefont {Watanabe}, \citenamefont {Taniguchi}, \citenamefont
  {Bernevig},\ and\ \citenamefont {Yazdani}}]{Nuckolls2020}%
  \BibitemOpen
  \bibfield  {author} {\bibinfo {author} {\bibfnamefont {K.~P.}\ \bibnamefont
  {Nuckolls}}, \bibinfo {author} {\bibfnamefont {M.}~\bibnamefont {Oh}},
  \bibinfo {author} {\bibfnamefont {D.}~\bibnamefont {Wong}}, \bibinfo {author}
  {\bibfnamefont {B.}~\bibnamefont {Lian}}, \bibinfo {author} {\bibfnamefont
  {K.}~\bibnamefont {Watanabe}}, \bibinfo {author} {\bibfnamefont
  {T.}~\bibnamefont {Taniguchi}}, \bibinfo {author} {\bibfnamefont {B.~A.}\
  \bibnamefont {Bernevig}}, \ and\ \bibinfo {author} {\bibfnamefont
  {A.}~\bibnamefont {Yazdani}},\ }\bibfield  {title} {\enquote {\bibinfo
  {title} {Strongly correlated {Chern} insulators in magic-angle twisted
  bilayer graphene},}\ }\href {\doibase 10.1038/s41586-020-3028-8} {\bibfield
  {journal} {\bibinfo  {journal} {Nature}\ }\textbf {\bibinfo {volume} {588}},\
  \bibinfo {pages} {610--615} (\bibinfo {year} {2020})}\BibitemShut {NoStop}%
\bibitem [{\citenamefont {Steinbrecher}\ \emph {et~al.}(2021)\citenamefont
  {Steinbrecher}, \citenamefont {van Weerdenburg}, \citenamefont {Walraven},
  \citenamefont {van Mullekom}, \citenamefont {Gerritsen}, \citenamefont
  {Natterer}, \citenamefont {Badrtdinov}, \citenamefont {Rudenko},
  \citenamefont {Mazurenko}, \citenamefont {Katsnelson}, \citenamefont {van~der
  Avoird}, \citenamefont {Groenenboom},\ and\ \citenamefont
  {Khajetoorians}}]{Steinbrecher2021}%
  \BibitemOpen
  \bibfield  {author} {\bibinfo {author} {\bibfnamefont {M.}~\bibnamefont
  {Steinbrecher}}, \bibinfo {author} {\bibfnamefont {W.~M.~J.}\ \bibnamefont
  {van Weerdenburg}}, \bibinfo {author} {\bibfnamefont {E.~F.}\ \bibnamefont
  {Walraven}}, \bibinfo {author} {\bibfnamefont {N.~P.~E.}\ \bibnamefont {van
  Mullekom}}, \bibinfo {author} {\bibfnamefont {J.~W.}\ \bibnamefont
  {Gerritsen}}, \bibinfo {author} {\bibfnamefont {F.~D.}\ \bibnamefont
  {Natterer}}, \bibinfo {author} {\bibfnamefont {D.~I.}\ \bibnamefont
  {Badrtdinov}}, \bibinfo {author} {\bibfnamefont {A.~N.}\ \bibnamefont
  {Rudenko}}, \bibinfo {author} {\bibfnamefont {V.~V.}\ \bibnamefont
  {Mazurenko}}, \bibinfo {author} {\bibfnamefont {M.~I.}\ \bibnamefont
  {Katsnelson}}, \bibinfo {author} {\bibfnamefont {A.}~\bibnamefont {van~der
  Avoird}}, \bibinfo {author} {\bibfnamefont {G.~C.}\ \bibnamefont
  {Groenenboom}}, \ and\ \bibinfo {author} {\bibfnamefont {A.~A.}\ \bibnamefont
  {Khajetoorians}},\ }\href@noop {} {\enquote {\bibinfo {title} {Quantifying
  the interplay between fine structure and geometry of an individual molecule
  on a surface},}\ } (\bibinfo {year} {2021}),\ \Eprint
  {http://arxiv.org/abs/2007.01928} {arXiv:2007.01928 [cond-mat.mes-hall]}
  \BibitemShut {NoStop}%
\bibitem [{\citenamefont {Moussy}, \citenamefont {Courtois},\ and\
  \citenamefont {Pannetier}(2001)}]{Moussy2001}%
  \BibitemOpen
  \bibfield  {author} {\bibinfo {author} {\bibfnamefont {N.}~\bibnamefont
  {Moussy}}, \bibinfo {author} {\bibfnamefont {H.}~\bibnamefont {Courtois}}, \
  and\ \bibinfo {author} {\bibfnamefont {B.}~\bibnamefont {Pannetier}},\
  }\bibfield  {title} {\enquote {\bibinfo {title} {A very low temperature
  scanning tunneling microscope for the local spectroscopy of mesoscopic
  structures},}\ }\href {\doibase 10.1063/1.1331328} {\bibfield  {journal}
  {\bibinfo  {journal} {Rev. Sci. Instrum.}\ }\textbf {\bibinfo {volume}
  {72}},\ \bibinfo {pages} {128--131} (\bibinfo {year} {2001})}\BibitemShut
  {NoStop}%
\bibitem [{\citenamefont {le~Sueur}\ and\ \citenamefont
  {Joyez}(2006)}]{leSueur2006}%
  \BibitemOpen
  \bibfield  {author} {\bibinfo {author} {\bibfnamefont {H.}~\bibnamefont
  {le~Sueur}}\ and\ \bibinfo {author} {\bibfnamefont {P.}~\bibnamefont
  {Joyez}},\ }\bibfield  {title} {\enquote {\bibinfo {title} {Room-temperature
  tunnel current amplifier and experimental setup for high resolution
  electronic spectroscopy in millikelvin scanning tunneling microscope
  experiments},}\ }\href {\doibase 10.1063/1.2400024} {\bibfield  {journal}
  {\bibinfo  {journal} {Rev. Sci. Instrum.}\ }\textbf {\bibinfo {volume}
  {77}},\ \bibinfo {pages} {123701} (\bibinfo {year} {2006})}\BibitemShut
  {NoStop}%
\bibitem [{\citenamefont {Kambara}\ \emph {et~al.}(2007)\citenamefont
  {Kambara}, \citenamefont {Matsui}, \citenamefont {Niimi},\ and\ \citenamefont
  {Fukuyama}}]{Kambara2007}%
  \BibitemOpen
  \bibfield  {author} {\bibinfo {author} {\bibfnamefont {H.}~\bibnamefont
  {Kambara}}, \bibinfo {author} {\bibfnamefont {T.}~\bibnamefont {Matsui}},
  \bibinfo {author} {\bibfnamefont {Y.}~\bibnamefont {Niimi}}, \ and\ \bibinfo
  {author} {\bibfnamefont {H.}~\bibnamefont {Fukuyama}},\ }\bibfield  {title}
  {\enquote {\bibinfo {title} {Construction of a versatile ultralow temperature
  scanning tunneling microscope},}\ }\href {\doibase 10.1063/1.2751095}
  {\bibfield  {journal} {\bibinfo  {journal} {Rev. Sci. Instrum.}\ }\textbf
  {\bibinfo {volume} {78}},\ \bibinfo {pages} {073703} (\bibinfo {year}
  {2007})}\BibitemShut {NoStop}%
\bibitem [{\citenamefont {Song}\ \emph
  {et~al.}(2010{\natexlab{b}})\citenamefont {Song}, \citenamefont {Otte},
  \citenamefont {Shvarts}, \citenamefont {Zhao}, \citenamefont {Kuk},
  \citenamefont {Blankenship}, \citenamefont {Band}, \citenamefont {Hess},\
  and\ \citenamefont {Stroscio}}]{Song2010}%
  \BibitemOpen
  \bibfield  {author} {\bibinfo {author} {\bibfnamefont {Y.~J.}\ \bibnamefont
  {Song}}, \bibinfo {author} {\bibfnamefont {A.~F.}\ \bibnamefont {Otte}},
  \bibinfo {author} {\bibfnamefont {V.}~\bibnamefont {Shvarts}}, \bibinfo
  {author} {\bibfnamefont {Z.}~\bibnamefont {Zhao}}, \bibinfo {author}
  {\bibfnamefont {Y.}~\bibnamefont {Kuk}}, \bibinfo {author} {\bibfnamefont
  {S.~R.}\ \bibnamefont {Blankenship}}, \bibinfo {author} {\bibfnamefont
  {A.}~\bibnamefont {Band}}, \bibinfo {author} {\bibfnamefont {F.~M.}\
  \bibnamefont {Hess}}, \ and\ \bibinfo {author} {\bibfnamefont {J.~A.}\
  \bibnamefont {Stroscio}},\ }\bibfield  {title} {\enquote {\bibinfo {title}
  {Invited review article: A 10 {mK} scanning probe microscopy facility},}\
  }\href {\doibase 10.1063/1.3520482} {\bibfield  {journal} {\bibinfo
  {journal} {Rev. Sci. Instrum.}\ }\textbf {\bibinfo {volume} {81}},\ \bibinfo
  {pages} {121101} (\bibinfo {year} {2010}{\natexlab{b}})}\BibitemShut
  {NoStop}%
\bibitem [{\citenamefont {Singh}\ \emph {et~al.}(2013)\citenamefont {Singh},
  \citenamefont {Enayat}, \citenamefont {White},\ and\ \citenamefont
  {Wahl}}]{Singh2013}%
  \BibitemOpen
  \bibfield  {author} {\bibinfo {author} {\bibfnamefont {U.~R.}\ \bibnamefont
  {Singh}}, \bibinfo {author} {\bibfnamefont {M.}~\bibnamefont {Enayat}},
  \bibinfo {author} {\bibfnamefont {S.~C.}\ \bibnamefont {White}}, \ and\
  \bibinfo {author} {\bibfnamefont {P.}~\bibnamefont {Wahl}},\ }\bibfield
  {title} {\enquote {\bibinfo {title} {Construction and performance of a
  dilution-refrigerator based spectroscopic-imaging scanning tunneling
  microscope},}\ }\href {\doibase 10.1063/1.4788941} {\bibfield  {journal}
  {\bibinfo  {journal} {Rev. Sci. Instrum.}\ }\textbf {\bibinfo {volume}
  {84}},\ \bibinfo {pages} {013708} (\bibinfo {year} {2013})}\BibitemShut
  {NoStop}%
\bibitem [{\citenamefont {Assig}\ \emph {et~al.}(2013)\citenamefont {Assig},
  \citenamefont {Etzkorn}, \citenamefont {Enders}, \citenamefont {Stiepany},
  \citenamefont {Ast},\ and\ \citenamefont {Kern}}]{Assig2013}%
  \BibitemOpen
  \bibfield  {author} {\bibinfo {author} {\bibfnamefont {M.}~\bibnamefont
  {Assig}}, \bibinfo {author} {\bibfnamefont {M.}~\bibnamefont {Etzkorn}},
  \bibinfo {author} {\bibfnamefont {A.}~\bibnamefont {Enders}}, \bibinfo
  {author} {\bibfnamefont {W.}~\bibnamefont {Stiepany}}, \bibinfo {author}
  {\bibfnamefont {C.~R.}\ \bibnamefont {Ast}}, \ and\ \bibinfo {author}
  {\bibfnamefont {K.}~\bibnamefont {Kern}},\ }\bibfield  {title} {\enquote
  {\bibinfo {title} {A 10 {mK} scanning tunneling microscope operating in ultra
  high vacuum and high magnetic fields},}\ }\href {\doibase 10.1063/1.4793793}
  {\bibfield  {journal} {\bibinfo  {journal} {Rev. Sci. Instrum.}\ }\textbf
  {\bibinfo {volume} {84}},\ \bibinfo {pages} {033903} (\bibinfo {year}
  {2013})}\BibitemShut {NoStop}%
\bibitem [{\citenamefont {Misra}\ \emph {et~al.}(2013)\citenamefont {Misra},
  \citenamefont {Zhou}, \citenamefont {Drozdov}, \citenamefont {Seo},
  \citenamefont {Urban}, \citenamefont {Gyenis}, \citenamefont {Kingsley},
  \citenamefont {Jones},\ and\ \citenamefont {Yazdani}}]{Misra2013}%
  \BibitemOpen
  \bibfield  {author} {\bibinfo {author} {\bibfnamefont {S.}~\bibnamefont
  {Misra}}, \bibinfo {author} {\bibfnamefont {B.~B.}\ \bibnamefont {Zhou}},
  \bibinfo {author} {\bibfnamefont {I.~K.}\ \bibnamefont {Drozdov}}, \bibinfo
  {author} {\bibfnamefont {J.}~\bibnamefont {Seo}}, \bibinfo {author}
  {\bibfnamefont {L.}~\bibnamefont {Urban}}, \bibinfo {author} {\bibfnamefont
  {A.}~\bibnamefont {Gyenis}}, \bibinfo {author} {\bibfnamefont {S.~C.~J.}\
  \bibnamefont {Kingsley}}, \bibinfo {author} {\bibfnamefont {H.}~\bibnamefont
  {Jones}}, \ and\ \bibinfo {author} {\bibfnamefont {A.}~\bibnamefont
  {Yazdani}},\ }\bibfield  {title} {\enquote {\bibinfo {title} {Design and
  performance of an ultra-high vacuum scanning tunneling microscope operating
  at dilution refrigerator temperatures and high magnetic fields},}\ }\href
  {\doibase 10.1063/1.4822271} {\bibfield  {journal} {\bibinfo  {journal} {Rev.
  Sci. Instrum.}\ }\textbf {\bibinfo {volume} {84}},\ \bibinfo {pages} {103903}
  (\bibinfo {year} {2013})}\BibitemShut {NoStop}%
\bibitem [{\citenamefont {Roychowdhury}\ \emph {et~al.}(2014)\citenamefont
  {Roychowdhury}, \citenamefont {Gubrud}, \citenamefont {Dana}, \citenamefont
  {Anderson}, \citenamefont {Lobb}, \citenamefont {Wellstood},\ and\
  \citenamefont {Dreyer}}]{Roychowdhury2014}%
  \BibitemOpen
  \bibfield  {author} {\bibinfo {author} {\bibfnamefont {A.}~\bibnamefont
  {Roychowdhury}}, \bibinfo {author} {\bibfnamefont {M.~A.}\ \bibnamefont
  {Gubrud}}, \bibinfo {author} {\bibfnamefont {R.}~\bibnamefont {Dana}},
  \bibinfo {author} {\bibfnamefont {J.~R.}\ \bibnamefont {Anderson}}, \bibinfo
  {author} {\bibfnamefont {C.~J.}\ \bibnamefont {Lobb}}, \bibinfo {author}
  {\bibfnamefont {F.~C.}\ \bibnamefont {Wellstood}}, \ and\ \bibinfo {author}
  {\bibfnamefont {M.}~\bibnamefont {Dreyer}},\ }\bibfield  {title} {\enquote
  {\bibinfo {title} {A 30 {mK}, 13.5 {T} scanning tunneling microscope with two
  independent tips},}\ }\href {\doibase 10.1063/1.4871056} {\bibfield
  {journal} {\bibinfo  {journal} {Rev. Sci. Instrum.}\ }\textbf {\bibinfo
  {volume} {85}},\ \bibinfo {pages} {043706} (\bibinfo {year}
  {2014})}\BibitemShut {NoStop}%
\bibitem [{\citenamefont {von Allw\"orden}\ \emph {et~al.}(2018)\citenamefont
  {von Allw\"orden}, \citenamefont {Eich}, \citenamefont {Knol}, \citenamefont
  {Hermenau}, \citenamefont {Sonntag}, \citenamefont {Gerritsen}, \citenamefont
  {Wegner},\ and\ \citenamefont {Khajetoorians}}]{vonAllworden2018}%
  \BibitemOpen
  \bibfield  {author} {\bibinfo {author} {\bibfnamefont {H.}~\bibnamefont {von
  Allw\"orden}}, \bibinfo {author} {\bibfnamefont {A.}~\bibnamefont {Eich}},
  \bibinfo {author} {\bibfnamefont {E.~J.}\ \bibnamefont {Knol}}, \bibinfo
  {author} {\bibfnamefont {J.}~\bibnamefont {Hermenau}}, \bibinfo {author}
  {\bibfnamefont {A.}~\bibnamefont {Sonntag}}, \bibinfo {author} {\bibfnamefont
  {J.~W.}\ \bibnamefont {Gerritsen}}, \bibinfo {author} {\bibfnamefont
  {D.}~\bibnamefont {Wegner}}, \ and\ \bibinfo {author} {\bibfnamefont {A.~A.}\
  \bibnamefont {Khajetoorians}},\ }\bibfield  {title} {\enquote {\bibinfo
  {title} {Design and performance of an ultra-high vacuum spin-polarized
  scanning tunneling microscope operating at 30 {mK} and in a vector magnetic
  field},}\ }\href {\doibase 10.1063/1.5020045} {\bibfield  {journal} {\bibinfo
   {journal} {Rev. Sci. Instrum.}\ }\textbf {\bibinfo {volume} {89}},\ \bibinfo
  {pages} {033902} (\bibinfo {year} {2018})}\BibitemShut {NoStop}%
\bibitem [{\citenamefont {Machida}, \citenamefont {Kohsaka},\ and\
  \citenamefont {Hanaguri}(2018)}]{Machida2018}%
  \BibitemOpen
  \bibfield  {author} {\bibinfo {author} {\bibfnamefont {T.}~\bibnamefont
  {Machida}}, \bibinfo {author} {\bibfnamefont {Y.}~\bibnamefont {Kohsaka}}, \
  and\ \bibinfo {author} {\bibfnamefont {T.}~\bibnamefont {Hanaguri}},\
  }\bibfield  {title} {\enquote {\bibinfo {title} {A scanning tunneling
  microscope for spectroscopic imaging below 90 {mK} in magnetic fields up to
  17.5 {T}},}\ }\href {\doibase 10.1063/1.5049619} {\bibfield  {journal}
  {\bibinfo  {journal} {Rev. Sci. Instrum.}\ }\textbf {\bibinfo {volume}
  {89}},\ \bibinfo {pages} {093707} (\bibinfo {year} {2018})}\BibitemShut
  {NoStop}%
\bibitem [{\citenamefont {Balashov}, \citenamefont {Meyer},\ and\ \citenamefont
  {Wulfhekel}(2018)}]{Balashov2018}%
  \BibitemOpen
  \bibfield  {author} {\bibinfo {author} {\bibfnamefont {T.}~\bibnamefont
  {Balashov}}, \bibinfo {author} {\bibfnamefont {M.}~\bibnamefont {Meyer}}, \
  and\ \bibinfo {author} {\bibfnamefont {W.}~\bibnamefont {Wulfhekel}},\
  }\bibfield  {title} {\enquote {\bibinfo {title} {A compact ultrahigh vacuum
  scanning tunneling microscope with dilution refrigeration},}\ }\href
  {\doibase 10.1063/1.5043636} {\bibfield  {journal} {\bibinfo  {journal} {Rev.
  Sci. Instrum.}\ }\textbf {\bibinfo {volume} {89}},\ \bibinfo {pages} {113707}
  (\bibinfo {year} {2018})}\BibitemShut {NoStop}%
\bibitem [{\citenamefont {Wong}\ \emph {et~al.}(2020)\citenamefont {Wong},
  \citenamefont {Jeon}, \citenamefont {Nuckolls}, \citenamefont {Oh},
  \citenamefont {Kingsley},\ and\ \citenamefont {Yazdani}}]{Wong2020}%
  \BibitemOpen
  \bibfield  {author} {\bibinfo {author} {\bibfnamefont {D.}~\bibnamefont
  {Wong}}, \bibinfo {author} {\bibfnamefont {S.}~\bibnamefont {Jeon}}, \bibinfo
  {author} {\bibfnamefont {K.~P.}\ \bibnamefont {Nuckolls}}, \bibinfo {author}
  {\bibfnamefont {M.}~\bibnamefont {Oh}}, \bibinfo {author} {\bibfnamefont
  {S.~C.~J.}\ \bibnamefont {Kingsley}}, \ and\ \bibinfo {author} {\bibfnamefont
  {A.}~\bibnamefont {Yazdani}},\ }\bibfield  {title} {\enquote {\bibinfo
  {title} {A modular ultra-high vacuum millikelvin scanning tunneling
  microscope},}\ }\href {\doibase 10.1063/1.5132872} {\bibfield  {journal}
  {\bibinfo  {journal} {Rev. Sci. Instrum.}\ }\textbf {\bibinfo {volume}
  {91}},\ \bibinfo {pages} {023703} (\bibinfo {year} {2020})}\BibitemShut
  {NoStop}%
\bibitem [{\citenamefont {Schwenk}\ \emph {et~al.}(2020)\citenamefont
  {Schwenk}, \citenamefont {Kim}, \citenamefont {Berwanger}, \citenamefont
  {Ghahari}, \citenamefont {Walkup}, \citenamefont {Slot}, \citenamefont {Le},
  \citenamefont {Cullen}, \citenamefont {Blankenship}, \citenamefont
  {Vranjkovic}, \citenamefont {Hug}, \citenamefont {Kuk}, \citenamefont
  {Giessibl},\ and\ \citenamefont {Stroscio}}]{Schwenk2020}%
  \BibitemOpen
  \bibfield  {author} {\bibinfo {author} {\bibfnamefont {J.}~\bibnamefont
  {Schwenk}}, \bibinfo {author} {\bibfnamefont {S.}~\bibnamefont {Kim}},
  \bibinfo {author} {\bibfnamefont {J.}~\bibnamefont {Berwanger}}, \bibinfo
  {author} {\bibfnamefont {F.}~\bibnamefont {Ghahari}}, \bibinfo {author}
  {\bibfnamefont {D.}~\bibnamefont {Walkup}}, \bibinfo {author} {\bibfnamefont
  {M.~R.}\ \bibnamefont {Slot}}, \bibinfo {author} {\bibfnamefont {S.~T.}\
  \bibnamefont {Le}}, \bibinfo {author} {\bibfnamefont {W.~G.}\ \bibnamefont
  {Cullen}}, \bibinfo {author} {\bibfnamefont {S.~R.}\ \bibnamefont
  {Blankenship}}, \bibinfo {author} {\bibfnamefont {S.}~\bibnamefont
  {Vranjkovic}}, \bibinfo {author} {\bibfnamefont {H.~J.}\ \bibnamefont {Hug}},
  \bibinfo {author} {\bibfnamefont {Y.}~\bibnamefont {Kuk}}, \bibinfo {author}
  {\bibfnamefont {F.~J.}\ \bibnamefont {Giessibl}}, \ and\ \bibinfo {author}
  {\bibfnamefont {J.~A.}\ \bibnamefont {Stroscio}},\ }\bibfield  {title}
  {\enquote {\bibinfo {title} {Achieving $\mu${eV} tunneling resolution in an
  \textit{in-operando} scanning tunneling microscopy, atomic force microscopy,
  and magnetotransport system for quantum materials research},}\ }\href
  {\doibase 10.1063/5.0005320} {\bibfield  {journal} {\bibinfo  {journal} {Rev.
  Sci. Instrum.}\ }\textbf {\bibinfo {volume} {91}},\ \bibinfo {pages} {071101}
  (\bibinfo {year} {2020})}\BibitemShut {NoStop}%
\bibitem [{\citenamefont {Pobell}(2007)}]{Pobell2007}%
  \BibitemOpen
  \bibfield  {author} {\bibinfo {author} {\bibfnamefont {F.}~\bibnamefont
  {Pobell}},\ }\href {\doibase 10.1007/978-3-540-46360-3} {\emph {\bibinfo
  {title} {{Matter and Methods at Low Temperatures}}}}\ (\bibinfo  {publisher}
  {Springer-Verlag Berlin Heidelberg},\ \bibinfo {year} {2007})\BibitemShut
  {NoStop}%
\bibitem [{\citenamefont {Voigtl{\"a}nder}\ \emph {et~al.}(2017)\citenamefont
  {Voigtl{\"a}nder}, \citenamefont {Coenen}, \citenamefont {Cherepanov},
  \citenamefont {Borgens}, \citenamefont {Duden},\ and\ \citenamefont
  {Tautz}}]{Voigtlander2017}%
  \BibitemOpen
  \bibfield  {author} {\bibinfo {author} {\bibfnamefont {B.}~\bibnamefont
  {Voigtl{\"a}nder}}, \bibinfo {author} {\bibfnamefont {P.}~\bibnamefont
  {Coenen}}, \bibinfo {author} {\bibfnamefont {V.}~\bibnamefont {Cherepanov}},
  \bibinfo {author} {\bibfnamefont {P.}~\bibnamefont {Borgens}}, \bibinfo
  {author} {\bibfnamefont {T.}~\bibnamefont {Duden}}, \ and\ \bibinfo {author}
  {\bibfnamefont {F.~S.}\ \bibnamefont {Tautz}},\ }\bibfield  {title} {\enquote
  {\bibinfo {title} {Low vibration laboratory with a single-stage vibration
  isolation for microscopy applications},}\ }\href {\doibase 10.1063/1.4975832}
  {\bibfield  {journal} {\bibinfo  {journal} {Rev. Sci. Instrum.}\ }\textbf
  {\bibinfo {volume} {88}},\ \bibinfo {pages} {023703} (\bibinfo {year}
  {2017})}\BibitemShut {NoStop}%
\bibitem [{VAB()}]{VAB}%
  \BibitemOpen
  \href@noop {} {}\bibinfo {note} {{VAb Vakuum-Anlagenbau GmbH,
  Marie-Curie-Str. 11, 25337 Elmshorn, Germany}}\BibitemShut {NoStop}%
\bibitem [{Foc()}]{Focus}%
  \BibitemOpen
  \href@noop {} {}\bibinfo {note} {{Model FDG 150, FOCUS GmbH, Neukirchner Str.
  2, 65510 Huenstetten, Germany}}\BibitemShut {NoStop}%
\bibitem [{Spe()}]{Specs}%
  \BibitemOpen
  \href@noop {} {}\bibinfo {note} {{Model ErLEED 150, SPECS Surface Nano
  Analysis GmbH, Voltastrasse 5, 13355 Berlin, Germany}}\BibitemShut {NoStop}%
\bibitem [{VAC()}]{VACOM}%
  \BibitemOpen
  \href@noop {} {}\bibinfo {note} {{VACOM Vakuum Komponenten \& Messtechnik
  GmbH, In den Br\"ucken\"ackern, 307751 Gro{\ss}l\"obichau,
  Germany}}\BibitemShut {NoStop}%
\bibitem [{All()}]{Allectra}%
  \BibitemOpen
  \href@noop {} {}\bibinfo {note} {{Model 242-SMAD50-C40-4, Allectra GmbH,
  Traubeneichenstr. 62-66, 16567 Sch\"onfliess, Germany}}\BibitemShut {NoStop}%
\bibitem [{SAE()}]{SAES}%
  \BibitemOpen
  \href@noop {} {}\bibinfo {note} {{Model NEXTorr Z100, SAES Getters S.p.A.,
  Viale Italia 77, 20045 Lainate, Italy}}\BibitemShut {NoStop}%
\bibitem [{Cry({\natexlab{a}})}]{Cryovac}%
  \BibitemOpen
  \href@noop {} {} ({\natexlab{a}}),\ \bibinfo {note} {{CryoVac GmbH \& Co KG,
  Heuserweg 14, 53842 Troisdorf, Germany}}\BibitemShut {NoStop}%
\bibitem [{Cry({\natexlab{b}})}]{Cryomagnetics}%
  \BibitemOpen
  \href@noop {} {} ({\natexlab{b}}),\ \bibinfo {note} {{Cryomagnetics, Inc.,
  1006 Alvin Weinberg Drive, Oak Ridge, TN 37830, USA}}\BibitemShut {NoStop}%
\bibitem [{\citenamefont {Haefer}(1989)}]{Haefer1989}%
  \BibitemOpen
  \bibfield  {author} {\bibinfo {author} {\bibfnamefont {R.}~\bibnamefont
  {Haefer}},\ }\href@noop {} {\emph {\bibinfo {title} {{Cryopumping: Theory and
  Practice}}}}\ (\bibinfo  {publisher} {Clarendon Press},\ \bibinfo {year}
  {1989})\BibitemShut {NoStop}%
\bibitem [{\citenamefont {Shvets}(1966)}]{Shvets1966}%
  \BibitemOpen
  \bibfield  {author} {\bibinfo {author} {\bibfnamefont {A.}~\bibnamefont
  {Shvets}},\ }\bibfield  {title} {\enquote {\bibinfo {title} {The use of
  liquid helium-3 to obtain temperatures down to 0.3k},}\ }\href {\doibase
  https://doi.org/10.1016/0011-2275(66)90130-5} {\bibfield  {journal} {\bibinfo
   {journal} {Cryogenics}\ }\textbf {\bibinfo {volume} {6}},\ \bibinfo {pages}
  {333--337} (\bibinfo {year} {1966})}\BibitemShut {NoStop}%
\bibitem [{\citenamefont {Torre}\ and\ \citenamefont
  {Chanin}(1985)}]{Torre1985}%
  \BibitemOpen
  \bibfield  {author} {\bibinfo {author} {\bibfnamefont {J.~P.}\ \bibnamefont
  {Torre}}\ and\ \bibinfo {author} {\bibfnamefont {G.}~\bibnamefont {Chanin}},\
  }\bibfield  {title} {\enquote {\bibinfo {title} {{Miniature liquid 3He
  refrigerator}},}\ }\href {\doibase 10.1063/1.1138350} {\bibfield  {journal}
  {\bibinfo  {journal} {Review of Scientific Instruments}\ }\textbf {\bibinfo
  {volume} {56}},\ \bibinfo {pages} {318--320} (\bibinfo {year}
  {1985})}\BibitemShut {NoStop}%
\bibitem [{Cha()}]{ChaseCryogenics}%
  \BibitemOpen
  \href@noop {} {}\bibinfo {note} {{Chase Research Cryogenics Ltd., Unit 2,
  Neepsend Industrial Estate, 80 Parkwood Rd, Neepsend, Sheffield S3 8AG,
  UK}}\BibitemShut {NoStop}%
\bibitem [{Ent()}]{Entropy}%
  \BibitemOpen
  \href@noop {} {}\bibinfo {note} {{Entropy GmbH, Gmunder Str. 37a, 81379
  Munich, Germany}}\BibitemShut {NoStop}%
\bibitem [{Lak()}]{Lakeshore}%
  \BibitemOpen
  \href@noop {} {}\bibinfo {note} {{Lake Shore Cryotronics, Inc., 550 Tressler
  Dr, Westerville, OH 43082, USA}}\BibitemShut {NoStop}%
\bibitem [{CMR()}]{CMR}%
  \BibitemOpen
  \href@noop {} {}\bibinfo {note} {{CMR-Direct, Willow House, 100 High Street,
  Somersham PE28 3EH, UK}}\BibitemShut {NoStop}%
\bibitem [{GVL()}]{GVL}%
  \BibitemOpen
  \href@noop {} {}\bibinfo {note} {{Type GVLZ141, GVL Cryoengineering Dr.
  George V. Lecomte GmbH, Aachener Strasse 89, 52223 Stolberg,
  Germany}}\BibitemShut {NoStop}%
\bibitem [{\citenamefont {Wilson}\ and\ \citenamefont
  {Timbie}(1999)}]{Wilson1999}%
  \BibitemOpen
  \bibfield  {author} {\bibinfo {author} {\bibfnamefont {G.~W.}\ \bibnamefont
  {Wilson}}\ and\ \bibinfo {author} {\bibfnamefont {P.~T.}\ \bibnamefont
  {Timbie}},\ }\bibfield  {title} {\enquote {\bibinfo {title} {{Construction
  techniques for adiabatic demagnetization refrigerators using ferric ammonium
  alum}},}\ }\href {\doibase http://dx.doi.org/10.1016/S0011-2275(99)00049-1}
  {\bibfield  {journal} {\bibinfo  {journal} {Cryogenics}\ }\textbf {\bibinfo
  {volume} {39}},\ \bibinfo {pages} {319--322} (\bibinfo {year}
  {1999})}\BibitemShut {NoStop}%
\bibitem [{\citenamefont {Jang}\ \emph {et~al.}(2015)\citenamefont {Jang},
  \citenamefont {Gruner}, \citenamefont {Steppke}, \citenamefont {Mitsumoto},
  \citenamefont {Geibel},\ and\ \citenamefont {Brando}}]{Jang2015}%
  \BibitemOpen
  \bibfield  {author} {\bibinfo {author} {\bibfnamefont {D.}~\bibnamefont
  {Jang}}, \bibinfo {author} {\bibfnamefont {T.}~\bibnamefont {Gruner}},
  \bibinfo {author} {\bibfnamefont {A.}~\bibnamefont {Steppke}}, \bibinfo
  {author} {\bibfnamefont {K.}~\bibnamefont {Mitsumoto}}, \bibinfo {author}
  {\bibfnamefont {C.}~\bibnamefont {Geibel}}, \ and\ \bibinfo {author}
  {\bibfnamefont {M.}~\bibnamefont {Brando}},\ }\bibfield  {title} {\enquote
  {\bibinfo {title} {{Large magnetocaloric effect and adiabatic demagnetization
  refrigeration with YbPt$_{2}$Sn}},}\ }\href {\doibase 10.1038/ncomms9680}
  {\bibfield  {journal} {\bibinfo  {journal} {Nature Communications}\ }\textbf
  {\bibinfo {volume} {6}},\ \bibinfo {pages} {8680} (\bibinfo {year}
  {2015})}\BibitemShut {NoStop}%
\bibitem [{\citenamefont {Wikus}, \citenamefont {Burghart{\'{i}}},\ and\
  \citenamefont {Figueroa-Feliciano}(2011)}]{Wikus2011}%
  \BibitemOpen
  \bibfield  {author} {\bibinfo {author} {\bibfnamefont {P.}~\bibnamefont
  {Wikus}}, \bibinfo {author} {\bibfnamefont {G.}~\bibnamefont
  {Burghart{\'{i}}}}, \ and\ \bibinfo {author} {\bibfnamefont {E.}~\bibnamefont
  {Figueroa-Feliciano}},\ }\bibfield  {title} {\enquote {\bibinfo {title}
  {{Optimum operating regimes of common paramagnetic refrigerants}},}\ }\href
  {\doibase 10.1016/j.cryogenics.2011.07.001} {\bibfield  {journal} {\bibinfo
  {journal} {Cryogenics}\ }\textbf {\bibinfo {volume} {51}},\ \bibinfo {pages}
  {555--558} (\bibinfo {year} {2011})}\BibitemShut {NoStop}%
\bibitem [{\citenamefont {Vilches}\ and\ \citenamefont
  {Wheatley}(1966)}]{Vilches1966}%
  \BibitemOpen
  \bibfield  {author} {\bibinfo {author} {\bibfnamefont {O.~E.}\ \bibnamefont
  {Vilches}}\ and\ \bibinfo {author} {\bibfnamefont {J.~C.}\ \bibnamefont
  {Wheatley}},\ }\bibfield  {title} {\enquote {\bibinfo {title} {{Measurements
  of the specific heats of three magnetic salts at low temperatures}},}\ }\href
  {\doibase 10.1103/PhysRev.148.509} {\bibfield  {journal} {\bibinfo  {journal}
  {Phys. Rev.}\ }\textbf {\bibinfo {volume} {148}},\ \bibinfo {pages}
  {509--516} (\bibinfo {year} {1966})}\BibitemShut {NoStop}%
\bibitem [{\citenamefont {Torre}\ and\ \citenamefont
  {Chanin}(1984)}]{Torre1984}%
  \BibitemOpen
  \bibfield  {author} {\bibinfo {author} {\bibfnamefont {J.~P.}\ \bibnamefont
  {Torre}}\ and\ \bibinfo {author} {\bibfnamefont {G.}~\bibnamefont {Chanin}},\
  }\bibfield  {title} {\enquote {\bibinfo {title} {{Heat switch for
  liquid-helium temperatures}},}\ }\href {\doibase 10.1063/1.1137726}
  {\bibfield  {journal} {\bibinfo  {journal} {Rev. Sci. Instrum.}\ }\textbf
  {\bibinfo {volume} {55}},\ \bibinfo {pages} {213--215} (\bibinfo {year}
  {1984})}\BibitemShut {NoStop}%
\bibitem [{\citenamefont {Schiffer}\ \emph {et~al.}(1994)\citenamefont
  {Schiffer}, \citenamefont {Ramirez}, \citenamefont {Huse},\ and\
  \citenamefont {Valentino}}]{Schiffer1994}%
  \BibitemOpen
  \bibfield  {author} {\bibinfo {author} {\bibfnamefont {P.}~\bibnamefont
  {Schiffer}}, \bibinfo {author} {\bibfnamefont {A.~P.}\ \bibnamefont
  {Ramirez}}, \bibinfo {author} {\bibfnamefont {D.~A.}\ \bibnamefont {Huse}}, \
  and\ \bibinfo {author} {\bibfnamefont {A.~J.}\ \bibnamefont {Valentino}},\
  }\bibfield  {title} {\enquote {\bibinfo {title} {{Investigation of the field
  induced antiferromagnetic phase transition in the frustrated magnet:
  Gadolinium gallium garnet}},}\ }\href {\doibase 10.1103/PhysRevLett.73.2500}
  {\bibfield  {journal} {\bibinfo  {journal} {Phys. Rev. Lett.}\ }\textbf
  {\bibinfo {volume} {73}},\ \bibinfo {pages} {2500--2503} (\bibinfo {year}
  {1994})}\BibitemShut {NoStop}%
\bibitem [{NF()}]{NF}%
  \BibitemOpen
  \href@noop {} {}\bibinfo {note} {{Model SA-606F2, NF Corporation, 6-3-20
  Tsunashima Higashi, Kohoku-ku, Yokohama 223-8508, Japan}}\BibitemShut
  {NoStop}%
\bibitem [{Nan()}]{Nanonis}%
  \BibitemOpen
  \href@noop {} {}\bibinfo {note} {{Nanonis SPM control system, SPECS Zurich
  GmbH, Technoparkstrasse 1, 8005 Zurich, Switzerland}}\BibitemShut {NoStop}%
\bibitem [{API()}]{API_bias}%
  \BibitemOpen
  \href@noop {} {}\bibinfo {note} {{Model 51-726-017, API Technologies Spectrum
  Control GmbH, Hansastrasse 6, 91126 Schwabach, Germany}}\BibitemShut
  {NoStop}%
\bibitem [{Fil()}]{Filter_piezo}%
  \BibitemOpen
  \href@noop {} {}\bibinfo {note} {{Tusonix / CTS 4106-001LF, Mouser part
  number 800-4106-001LF, 1000 N. Main Street, Mansfield, TX 76063,
  USA}}\BibitemShut {NoStop}%
\bibitem [{Fem()}]{Femto}%
  \BibitemOpen
  \href@noop {} {}\bibinfo {note} {{Model DLPCA-200, FEMTO Messtechnik GmbH,
  Klosterstr. 64, 10179 Berlin, Germany}}\BibitemShut {NoStop}%
\end{thebibliography}%

\end{document}